 \documentclass{emulateapj} 
 \usepackage{amsmath}

\providecommand{\e}[1]{\ensuremath{\times 10^{#1}}}
\newcommand{\Nvsg}{N_{\rm{VSG}}^{\tau=1}}
\newcommand{\sub}[1]{_{\rm{#1}}}

\begin{document}

\shorttitle{MRI by FUV Ionization} \shortauthors{Perez-Becker \& Chiang}

\title{SURFACE LAYER ACCRETION \\IN CONVENTIONAL AND TRANSITIONAL DISKS \\DRIVEN BY FAR-ULTRAVIOLET IONIZATION}

\author{Daniel Perez-Becker\altaffilmark{}} \affil{Department of Physics, University of California, Berkeley, CA 94720, USA}
\email{Electronic address: perez-becker@berkeley.edu}
\and

\author{Eugene Chiang\altaffilmark{}} \affil{Departments of Astronomy and Earth and Planetary Science, University of California, Berkeley, CA 94720, USA} 

\begin{abstract}
  Whether protoplanetary disks accrete at observationally significant
  rates by the magnetorotational instability (MRI) depends on how well
  ionized they are. Disk surface layers ionized by stellar X-rays are
  susceptible to charge neutralization by small condensates, ranging
  from $\sim$0.01-$\mu$m-sized grains to angstrom-sized
  polycyclic aromatic hydrocarbons (PAHs). Ion densities in
  X-ray-irradiated surfaces are so low that ambipolar diffusion
  weakens the MRI.  Here we show that ionization by stellar
  far-ultraviolet (FUV) radiation enables full-blown MRI turbulence in
  disk surface layers. Far-UV ionization of atomic carbon and sulfur
  produces a plasma so dense that it is immune to ion recombination on
  grains and PAHs.  The FUV-ionized layer, of thickness 0.01--0.1
  g/cm$^2$, behaves in the ideal magnetohydrodynamic limit and can
  accrete at observationally significant rates at radii $\gtrsim
  1$--10 AU. Surface layer accretion driven by FUV ionization can
  reproduce the trend of increasing accretion rate with
  increasing hole size seen in transitional disks. At radii $\lesssim$
  1--10 AU, FUV-ionized surface layers cannot sustain the accretion
  rates generated at larger distance, and unless turbulent mixing of plasma
  can thicken the MRI-active layer,
  an additional means of
  transport is needed.  In the case of transitional disks, it could be
  provided by planets.
\end{abstract}


\section{INTRODUCTION} \label{sec_intro}

T Tauri stars, Herbig Ae/Be stars, and young brown dwarfs accrete
\citep[e.g.,][]{Hartmann:2006p9080}. Perhaps the most direct evidence
for accretion comes from observations of radiation at blue to
ultraviolet wavelengths, emitted in excess of the Wien tail of the
stellar photosphere. The ultraviolet excess arises in large part from
gas that is shock heated upon free-falling onto the stellar surface
\citep{Calvet:1998p9042, JohnsKrull:2000p9448, Calvet:2004p9453}. From
these observations, and others of optical emission lines (e.g.,
\citealt{Hartmann:1994p10692}; \citealt{Herczeg:2008p11290}),
it is inferred that young solar mass stars
accrete at rates $\dot{M} \approx 10^{-9}$--$10^{-7} M\sub{\odot}$
yr$^{-1}$ (e.g., \citealt{Muzerolle:2005p9975}). Any individual system may
exhibit order-unity variations in $\dot{M}$ with time \citep{Eisner:2010p10573}.

Strictly speaking, these observations imply only that gas in the
immediate stellar vicinity---i.e., within a few stellar radii, in and
around the star's magnetosphere---is accreting.  Gas and dust orbit at
much greater stellocentric distances, residing in disks that in many
cases extend continuously from $\sim$0.1 to $\sim$100 AU (e.g.,
\citealt{Watson:2007p6076}), but there is no direct indication that
any of this material is actually accreting (but see
\citealt{Hughes:2011p10580} for an attempt at measuring non-Keplerian
motions of circumstellar gas).  Nevertheless the standard assumption
is that infalling gas near the star was transported there through the
disk, traveling decades in radius over the system age of $t_{\rm age}
\sim 10^6$ yr.  This assumption is at least consistent with the fact
that disk masses at large radii---as inferred from continuum dust
emission at radio wavelengths (e.g.,
\citealt{Andrews:2005p10575})---are similar to $\dot{M} t_{\rm age}$.

One of the most intensively studied mechanisms for transporting
material is turbulence driven by the magnetorotational instability
(MRI; for a review, see \citealt{Balbus:2009p5490}). The MRI requires that gas be ionized enough to couple dynamically to disk magnetic
fields. \citet{Gammie:1996p3339} appreciated that for 
protoplanetary disks, which are predominantly neutral, the MRI
may not operate everywhere. Over a wide range of
radii at the disk midplane, the magnetic Reynolds number
\begin{equation}\label{Re}
Re \equiv \frac{c_{\rm s}h}{D} \approx 1 \left( \frac{x_{\rm
      e}}{10^{-13}} \right) 
\left( \frac{T}{100\, {\rm K}} \right)^{1/2} 
\left( \frac{a}{{\rm AU}} \right)^{3/2} 
\end{equation}
is so low that MRI turbulence cannot be sustained against Ohmic
dissipation. The magnetic Reynolds number compares the strength of
magnetic induction (flux freezing) to Ohmic dissipation. Contemporary
numerical simulations suggest that the MRI requires $Re > Re^\ast
\approx 10^2$--$10^4$, the threshold depending on the initial field
geometry (e.g., \citealt{Fleming:2000p6431}).  In equation (\ref{Re}),
$a$ is the disk radius; $T$ is the gas temperature; $c_{\rm s}$ is the
gas sound speed; $h = c_{\rm s}/\Omega$ is the gas scale height;
$\Omega$ is the Kepler orbital frequency; $D = 234 \, (T/{\rm
  K})^{1/2} \, x_{\rm e}^{-1}$ cm$^2$ s$^{-1}$ is the magnetic
diffusivity; and $x_{\rm e}$ is the fractional abundance of electrons
by number.

\citet{Gammie:1996p3339} proposed further that MRI-dead midplane gas
is encased by MRI-active surface layers which satisfy the Reynolds
number criterion because they are ionized sufficiently by interstellar
cosmic rays. He estimated that the MRI-active layer extends down to a
surface density $\Sigma^\ast \approx 100$ g/cm$^2$---essentially the
stopping column for GeV-energy cosmic rays \citep{Umebayashi:1981p7866}. For
comparison, the total surface density at $a = 1$ AU in the
minimum-mass solar nebula is about 2200 g/cm$^2$ (e.g., \citealt{Chiang:2010p10583}). Subsequent work, discussed below, has shown that a fully MRI-active surface density of $\Sigma^\ast \approx 100$ g/cm$^2$ is
a gross overestimate.

\citet{Glassgold:1997p2130} pointed out that X-rays emitted by active
stellar coronae (e.g., \citealt{Preibisch:2005p6831}) are more
important than cosmic rays in determining the ionization fraction, in
part because disk surface layers are likely shielded from cosmic rays
by magnetized stellar outflows. Such shielding is almost certainly
present, as its effects are appreciable even for the comparatively feeble
Solar wind. At cosmic ray energies of 0.1--1 GeV, the present-day Solar
wind lowers the Galactic cosmic ray flux at Earth by roughly an order
of magnitude compared to estimated fluxes outside the heliosphere
(see Figure 1 of \citealt{Reedy:1987p11439}). 
Moreover, the anti-correlation between the terrestrial cosmic ray flux
and Solar activity has been clearly observed over many Solar cycles
\citep{Svensmark:1998p11442}.
By comparison with the Solar wind, pre-main-sequence winds have mass
loss rates that are five orders of magnitude larger, and thus we
expect the shielding of disk surface layers from Galactic cosmic rays
to be essentially complete for young stellar systems.
Nevertheless cosmic rays might still leak in from the
``side,'' entering the disk edge-on from the outside.  The impact of
``sideways'' cosmic rays is difficult to assess because whether cosmic
rays can penetrate radially inward depends on the radial distribution
of matter and on the magnetic field distribution, both of which are
unknown.  Cosmic rays might spiral toward the disk along interstellar
field lines only to be mirrored back. If we assume that this does not
happen, and are willing to be guided by the surface density profile of
the minimum-mass solar nebula, we estimate that ionization by sideways
cosmic rays might be significant on the outskirts of the disk at $a
\gtrsim 30$ AU.
Under these generous assumptions, cosmic rays can sustain MRI
turbulence in the entire outer disk---from the disk surface to the
midplane, at $a \gtrsim 30$ AU. We will return briefly to this scenario in
Section \ref{discussion}.

\citet{Bai:2009p3370}, \citet{Turner:2007p3503}, and
\citet{Turner:2010p6693} studied how disk surface layers could be made
MRI active by stellar X-rays.  They found $\Sigma^\ast \sim 1$--30
g/cm$^2$---the range over which photons of energy 1--10
keV are stopped. In their models, the precise extent of the
MRI-turbulent region depended on the abundance of charge-neutralizing
grains. X-ray ionized layers, though less deep than the layer
envisioned by \citet{Gammie:1996p3339}, were considered by \citet{Turner:2010p6693} to be capable of transporting mass at observationally
significant rates at $a = 1$ AU.

All of the aforementioned studies of layered accretion assumed that
Ohmic dissipation limits $\Sigma^\ast$, as gauged by $Re$ or the
related Elsasser number \citep{Turner:2007p3503}. This assumption,
widespread in the literature, overestimates $\Sigma^\ast$ and the
vigor of MRI turbulence in X-ray ionized surface layers.  In a recent
attempt to estimate $\Sigma^\ast$, \citeauthor{PerezBecker:2011p9801}
(2011, hereafter PBC11) employed an X-ray ionization model similar to
those used by previous workers, but accounted for two additional
factors.  First, following \citet{Chiang:2007p3804}, they tested
whether ambipolar diffusion---the decoupling of neutral matter from
plasma---limits $\Sigma^\ast$.  They gauged the importance of
ambipolar diffusion by computing at a given depth the number of times
a neutral H$_2$ molecule collides with charged species in a dynamical time
$\Omega^{-1}$:
\begin{equation}\label{Am}
Am \equiv \frac{n_{\rm charge} \beta_{\rm in}}{\Omega} \approx 1 \left( \frac{x_{\rm charge}}{10^{-8}} \right) \left( \frac{n_{\rm tot}}{10^{10}\, {\rm cm}^{-3}} \right) \left( \frac{a}{\rm AU} \right)^{3/2} \,.
\end{equation}
Here $n_{\rm charge} \equiv x_{\rm charge}n_{\rm tot}$ is the total number
density of singly charged species (assuming multiply charged species
are negligible); $x_{\rm charge}$ is the fractional charge density; $n_{\rm tot}$
is the number density of hydrogen nuclei (mostly in the form of
neutral H$_2$);
and $\beta_{\rm in} \approx 2 \times 10^{-9}$ cm$^3$
s$^{-1}$ is the collisional rate coefficient for singly charged species
to share their
momentum with neutrals (\citealt{Draine:1983p6430};
\citealt{Millar:1997p10585}).  Whereas $Re$ measures how well plasma
is tied to magnetic fields, $Am$ assesses how well neutral gas is
coupled to plasma. No matter how large $Re$ may be, unless $Am$ is
also large, the magnetic stresses felt by the plasma will not be effectively
communicated to the bulk of the disk gas, which is composed
overwhelmingly of neutral H$_2$ \citep{Blaes:1994p10586,Kunz:2004p11444,Desch:2004p11453,Wardle:2007p7104}.
Numerical simulations by \citeauthor{Hawley:1998p5481} (1998, hereafter HS)
suggested that unless $Am \gtrsim Am^\ast \approx 10^2$, MRI
turbulence would not be sustained in neutral gas. Ambipolar diffusion
is of especial concern in disk surface layers because $Am$ depends on
the absolute number density of charged particles, and not just the
relative density as in $Re$.  At a fixed ionization fraction, the
rapidly decreasing gas density with increasing height above the
midplane renders the MRI increasingly susceptible to ambipolar
diffusion.

The second factor considered by PBC11 concerned polycyclic
aromatic hydrocarbons (PAHs). Like grains, these macromolecules act
as sites of ion recombination. Their abundances in disk surface layers
were constrained by PBC11 using {\it Spitzer} satellite observations
of PAH emission lines, detected in a large fraction of Herbig Ae stars
surveyed \citep{Geers:2006p3349}.\footnote{Some workers wish to
  ignore PAHs when studying X-ray-driven chemistry on the grounds that
  they are rarely detected in T Tauri stars. It is hard for us to
  follow this argument. Herbig Ae disks commonly evince PAHs and
  exhibit accretion rates similar to if not higher than those of T
  Tauri disks. Treating the problem of T Tauri disk accretion separate
  from the problem of Herbig Ae disk accretion seems unjustified. As
  discussed by \citet{Geers:2006p3349}, PAHs usually go
  undetected in T Tauri stars not because they are absent but
  because they fluoresce less luminously in the weaker ultraviolet fields of
  their host stars.}
Disk PAHs were found to reduce $Am$ and $Re$ by factors $\gtrsim 10$
(PBC11).  For typical stellar X-ray spectra, $Am$ reached maximum
values on the order of unity at $\Sigma \sim 1$ g/cm$^2$.  Such
maximum values were attained only for the lowest plausible PAH
abundances. Comparing $\max Am \sim 1$ with HS's determination that
$Am^\ast \sim 10^2$, PBC11 concluded that ambipolar diffusion, abetted by PAHs,
reduced accretion rates in X-ray ionized surface layers to values
too small compared to observations.

The conclusion of PBC11 has since been tempered by two new developments.
First, PBC11 underestimated $Am$ because they omitted the contribution
from momentum-coupling collisions between H$_2$ and charged PAHs.
This neglect is potentially significant because
charged PAHs are
as well coupled to the magnetic field as are atomic/molecular ions (HCO$^+$
and metal ions like Mg$^+$ in their simple model).
They argued that because
the abundance of charged PAHs is less than that of molecular ions, $Am$
should not be significantly increased. This statement applies for the
lowest PAH abundance considered in PBC11 of $x\sub{PAH} \sim 10^{-11}$ (here
measured per H), but
\citet{Bai:2011p0001} and \citet{Mohanty:2011p0001} have pointed out that PAHs can dominate the charge budget
at higher $x\sub{PAH}$. 
In this high-PAH limit, although
PAHs decrease the free electron abundance and therefore $Re$, they can
boost $Am$---and in principle $\dot{M}$---in comparison to the low-PAH
case.
In this paper we correct for this effect by taking
\begin{equation} \label{ncharge}
n\sub{charge}=n\sub{i}+n\sub{PAH(Z=-1)}+n\sub{PAH(Z=1)}
\end{equation}
when computing $Am$, where the various densities
refer to atomic/molecular ions, PAHs with a single negative charge,
and PAHs with a single positive charge, respectively.

In another recent development, \citeauthor{Bai:2011p11460} (2011;
hereafter BS11) undertook new simulations of the MRI with ambipolar
diffusion to
update those of HS. They confirmed the result of HS that when $Am
\gtrsim 10$--$10^2$, neutrals and ions behave essentially as one fluid with
a Shakura-Sunyaev transport parameter $\alpha$ that can be as high as
$\sim$0.1--0.5.
Their new result is that MRI turbulence can still be sustained
in the neutrals for $Am \lesssim 1$, albeit in a weakened
form with $\max \alpha$ decreasing with decreasing $Am$.
When $Am \sim 1 (0.1)$, BS11 find that $\max \alpha \approx 0.01 (0.0007)$.
Whether $\alpha$ attains its maximum value for a given $Am$
depends on the assumed strength and geometry of the background 
magnetic field.

In this paper we consider a third, often neglected source of
ionization: far ultraviolet (FUV $\equiv$ photon energies
between $\sim$6 and 13.6 eV) radiation emitted by the central star. As
measured with the {\it Hubble Space Telescope} and {\it FUSE}, typical
FUV luminosities from young stars are $\gtrsim 10^{30}$ erg/s, much of
it in atomic lines (e.g., \citealt{Bergin:2007p9443}). This radiation
originates from the stellar accretion shock \citep{Calvet:1998p9042,
  JohnsKrull:2000p9448, Calvet:2004p9453} and from the active stellar
chromosphere \citep{Alexander:2005p10590}.
Photoionization/photodissociation regions generated by FUV radiation
have been studied extensively (e.g., \citealt{Gorti:2008p2645}, and
references therein; \citealt{Tielens:1985p7746}), but their
implications for the MRI have not been well quantified.
\citet{Semenov:2004p10917} carried out detailed calculations of the
disk ionization fraction due to X-ray and FUV radiation in order to
estimate the extent of the MRI-active layer, but they did not consider
the charge-adsorbing effects of PAHs or ambipolar diffusion.  We seek
here to give a more comprehensive treatment, including new estimates
of $\dot{M}$ driven by FUV ionization, made possible in part by the
powerful new results of BS11.

High in the disk atmosphere, molecules are photodissociated and
elements take their atomic form. We focus on FUV photons (having
energies less than the H Lyman limit) because they are not stopped by
atomic hydrogen and as such penetrate more deeply into the disk than
do Lyman continuum photons.  Far-UV photons ionize trace atoms. Some
elements, when nearly fully ionized, may be so abundant that the
criteria for MRI accretion are easily satisfied.  Carbon, for example,
has a first ionization energy of 11.26 eV and a cosmic number
abundance relative to hydrogen of approximately $2.9 \times 10^{-4}$
(\citealt{Lodders:2003p5309}, her Table 2).  The Str\"omgren layer of
CII generated by FUV radiation is characterized by ionization
fractions---and, by extension, values of $Am$ and $Re$---up to $10^5$
times larger than those reported by PBC11 for X-ray ionized layers of
H$_2$.  Thus ambipolar diffusion may not pose the same threat in
FUV-ionized layers that it does in X-ray ionized layers. Moreover,
electron densities in FUV-ionized layers may be so large that they 
are little impacted by PAHs or other small grains.

Of the various elements that can be ionized by FUV radiation, we model
in this paper only carbon and sulfur. Their cosmic abundances are
relatively high, and they are among the least likely elements to be
depleted onto grains (as evidenced in the diffuse interstellar medium;
e.g., \citealt{Jenkins:2009p6986}; \citealt{Savage:1996p10591}).  Line
emission from ionized carbon ([CII] 158 $\mu$m) has been detected by
\textit{Herschel} in protoplanetary disks
(\citealt{Pinte:2010p8649,Sturm:2010p8660}).  The observed line fluxes
can be reproduced by models that assume a cosmic gas phase abundance
of carbon \citep{Woitke:2010p8552}, although our own analysis shows
that lower abundances are also possible because the [CII]-emitting
layer is optically thick to its own emission.  The abundance of sulfur
has not been so constrained because no lines from sulfur have been
unambiguously detected in disks
(\citealt{Lahuis:2007p10598,Watson:2007p6076,Gorti:2008p2645}). Nevertheless
\citet{Meijerink:2008p8626} found that a sulfur abundance of up to $7
\times 10^{-6}$ per H, or $\sim$2/5 that of cosmic
\citep{Lodders:2003p5309}, could be reconciled with the non-detection
of the [SI] 25 $\mu$m line. The true upper limit on the sulfur
abundance may be even higher because \citet{Meijerink:2008p8626} did
not account for FUV ionization of SI.

Our paper is structured as follows. The equations governing the
ionization balance of carbon and sulfur in disk surface layers are
presented in Section \ref{model}. Account is made of PAHs; sub-micron
sized grains which can attenuate FUV radiation; and of H$_2$ which can
absorb photons having energies $> 11.2$ eV. Results for the carbon and
sulfur ionization fronts, and the values for $Am$, $Re$, and
$\Sigma^\ast$ they imply, are given in Section \ref{results}. 
The impact of the Hall effect (e.g., \citealt{Wardle:2011p11473}) on FUV-ionized
layers is also considered in Section \ref{results}.
A summary
is supplied in Section \ref{discussion}, together with estimates of
$\dot{M}$ as a function of disk radius, as driven by FUV radiation, X-rays, or cosmic rays.  We compare our results to stellar accretion rates
observed in conventional T Tauri disks and transitional disks with
inner optically thin holes.

\section{MODEL FOR FUV IONIZATION} \label{model}

At a given radius $a$ from a star of mass $M=1 M_\sun$, we compute the
equilibrium abundances of H$_2$, HI, CI, CII, SI, and SII as a function
of the vertical column penetrated by FUV radiation.  The key equations
governing the ratios of H$_2$:H, CI:CII, and SI:SII are, qualitatively, 

\begin{eqnarray}
{\rm Rate} & {\rm \,\, of} {\rm \,\, dissociation \,\, of \,\, }{\rm H}_2 {\rm \,\, by \,\,FUV \,\, photons}  \nonumber \\ = &{\rm Rate} {\rm \,\, of}  {\rm  \,\, formation \,\, of \,\,} {\rm H}_2 {\rm \,\, on \,\, grains} \, \,\label{quality_diss}
\end{eqnarray}

\begin{eqnarray} 
{\rm Rate} & {\rm\,\, of} {\rm \,\, ionization \,\, of \,\, }{\rm CI} {\rm \,\, by \,\,FUV \,\, photons} \nonumber \\  = & \!\!\!\! {\rm Rate} {\rm \,\, of}  {\rm  \,\, recombination \,\, of \,\,} {\rm CII} \nonumber \\ & {\rm \,\, with \,\, electrons, \,\, grains, \,\, and \,\, PAHs} \label{quality_carbon}
\end{eqnarray}

\noindent and

\begin{eqnarray} 
{\rm Rate} & {\rm\,\, of} {\rm \,\, ionization \,\, of \,\, }{\rm SI} {\rm \,\, by \,\,FUV \,\, photons} \nonumber \\  = & \!\!\!\! {\rm Rate} {\rm \,\, of}  {\rm  \,\, recombination \,\, of \,\,} {\rm SII} \nonumber \\ & {\rm \,\, with \,\, electrons, \,\, grains, \,\, and \,\, PAHs}\,. \label{quality_sulfur}
\end{eqnarray}

For our standard model we take $a = 3$ AU. This parameter and others
will be varied to explore their impact on the MRI-active column $\Sigma^\ast$.

\subsection{FUV Luminosity}\label{fuv}

Two ingredients of our model, H$_2$ and C, are
dissociated and ionized, respectively, by FUV photons having
practically the same wavelength range: $\lambda \approx 912$--$1109
\AA$.  A third ingredient, S, is ionized by photons having $\lambda <
1198 \AA$.  In the combined wavelength interval $\lambda \approx
912$--$1198 \AA$ we take the stellar luminosity of our standard model
to be $L_{\rm FUV} = 10^{30}$ erg/s, in continuum photons.  We take
$L_{\rm FUV}$ to be distributed uniformly over this wavelength interval
so that approximately $(2/3) L_{\rm FUV}$ can be absorbed by H$_2$, C,
and S, while the remaining $(1/3) L_{\rm FUV}$ can be absorbed by S
but not H$_2$ or C.

Our choice for the continuum luminosity is compatible with {\it FUSE}
observations of T Tauri stars (\citealt{Bergin:2003p9351}, see their Figure 1;
admittedly their spectra are extrapolated at $\lambda < 950 \AA$ for
TW Hydra and $\lambda < 1150 \AA$ for BP Tau). Our standard $L_{\rm
  FUV}$ underestimates the true ionizing luminosity because we neglect
commonly observed FUV emission lines (but note that our wavelength
range does not include the powerful H Lyman-alpha line at
$1216\AA$). It also neglects the large contribution
from the hotter photospheres of more massive stars (e.g.,
\citealt{MartinZaidi:2008p10601}). We make some account for these effects
by varying $L_{\rm FUV}$ up to $10^{32}$ erg/s in our parameter survey
(Section \ref{sec_sensitivity}).

\subsection{Total Gas Columns and Densities}
We present our results as a function of the total vertical hydrogen
column density $N\sub{tot} = N\sub{H}+ 2 N\sub{H_2}$, measured
perpendicular to and toward the disk midplane. An equivalent measure
is the mass surface density $\Sigma \equiv N\sub{tot} m\sub{H}$, where
$m\sub{H} = 1.7\e{-24}$ g is the mass of the hydrogen atom.
The local number density of hydrogen nuclei is approximated by
\begin{equation}
	n\sub{tot} \approx N\sub{tot}/h\,.
\end{equation}

Stellar radiation enters the disk at a grazing angle $\theta \sim 3h/a$
measured from the flared disk surface (e.g., \citealt{Chiang:2001p5086}).
We assume in this work that photons travel in straight paths, i.e.,
we neglect scattering of FUV radiation into directions other than that
of the original beam entering the disk. Thus
a beam of radiation that penetrates a vertical column $N\sub{tot}$
has traversed a larger column parallel to the incident beam direction
of $\sim$$N\sub{tot}/\theta$, and is attenuated according to the latter
quantity, not the former. To order of magnitude, $\theta \sim 0.3$.

\subsection{Gas Temperature} \label{gas_temp}
Gas temperatures in disk surface layers are set by a host of heating
and cooling processes. Models in the literature account differently
for these processes and yield temperatures which disagree. At a
vertical column of $N\sub{tot} \sim 10^{22}$ cm$^{-2}$---roughly where
the deepest of the fronts we compute, the SI/SII ionization front, 
may lie---\citeauthor{Glassgold:2007p9153} (2007, hereafter GNI07)
reported a gas temperature of $T \approx 60$ K at $a \approx 5$--10 AU
(see their Figure 2). By comparison, \citeauthor{Gorti:2008p2645} (2008,
hereafter GH08) found that at $a = 8$ AU, the S ionization front
occurs at $T \approx 600$ K (see their Figures 1 and 5).  One reason
for this difference seems to be that GNI07 omitted photoelectric
heating from PAHs, which dominates thermal balance according to
GH08. But the PAH abundance assumed by GH08 seems too
high, exceeding by about an order of magnitude the highest plausible
abundance inferred by PBC11. Reducing the heating rate from PAHs by
more than a factor of 10 from its value in GH08 (see their Figure 4) 
would imply that X-ray heating dominates thermal balance---as it tends to do by default in GNI07.

In this paper we do not solve the thermal balance equations, but
instead approximate the disk surface layer as vertically
isothermal at a temperature drawn from GNI07. This is similar to
the choice we made in PBC11, except that there we were interested
in X-ray ionized columns $N\sub{tot} \gtrsim 10^{22}$ cm$^{-2}$, whereas here
we are interested in FUV-ionized columns $N\sub{tot} \lesssim 10^{22}$ cm$^{-2}$.
As $N\sub{tot}$ decreases from $10^{22}$ to $10^{21}$ cm$^{-2}$,
gas temperatures in GNI07 rise by factors of $\sim$2--3.
At still higher altitudes for which $N\sub{tot} < 10^{21}$ cm$^{-2}$, temperatures rise steeply by an order of magnitude
or more. As a simple compromise we multiply the temperature profile of PBC11 by 
a factor of 3 but otherwise keep the same scaling:

\begin{equation}
T \approx 240 \left( \frac{a}{3 \, {\rm AU}} \right)^{-3/7} \, {\rm K} \,.
\end{equation}
Our assumption that the gas is vertically isothermal ignores the steep
temperature gradient at $N\sub{tot} < 10^{21}$
cm$^{-2}$ where dust-gas cooling is not effective.  This neglect
should not be serious as the amount of mass contained in these
uppermost layers is small compared to the mass in the MRI-active
layer, which we will find is concentrated near the C and S ionization
fronts, located within $N\sub{tot} \sim 10^{21}$--$10^{23}$ cm$^{-2}$.  Moreover,
the locations of the fronts are not especially sensitive to
temperature, as we have verified with a few test cases.

\subsection{Very Small Grains (VSGs)}\label{vsg_param}

Very small grains (VSGs) having sizes of order 0.01 $\mu$m can
attenuate FUV radiation.  We parameterize this absorption with the
variable $\Nvsg$, the total gas column for which VSGs present unit
optical depth in the FUV.\footnote{Equivalent to the variable
  $\sigma_{\rm H}$ used by GH08.}
We consider $\Nvsg$ between 10$^{21}$ and 10$^{24}$ cm$^{-2}$, taking
as our standard value $\Nvsg = 10^{22}$ cm$^{-2}$. In the diffuse
interstellar medium, $\Nvsg \approx 10^{21}$ cm$^{-2}$; its value in
disk surface layers may be much higher---by 1--3 orders of magnitude
as judged by disk infrared spectral energy distributions (see
Table 3 of PBC11 and references therein)---because of grain growth and
sedimentation.

Colliding with ions and faster moving electrons, VSGs carry on average
a net negative charge. They therefore act as sites of ion
recombination.  To account for such recombination, we model
VSGs as conducting spheres of radius $s\sub{VSG}=0.01$ $\mu$m. Their
number abundance relative to hydrogen is
\begin{equation} x_{\rm{VSG}}\equiv n_{\rm{VSG}}/n_{\rm{tot}}=\frac{1}{\Nvsg
    \pi s\sub{VSG}^2 } \,. \label{vsg_abun}
\end{equation}
In PBC11, we solved for the detailed charge distributions of
grains using a set of recurrence equations. Solving the recurrence
equations to obtain self-consistent solutions for the electron
density, the ion density, and the total charge carried by grains/PAHs
was necessary because in the lightly ionized
regions irradiated by X-rays, the charge carried
by PAHs could be significant compared to the amount of charge
in free electrons. By comparison, FUV ionized regions are simpler to treat
because free electrons are orders of magnitude more abundant
than grains/PAHs. Thus our analysis here is simplified:
we assume that every grain has
an equilibrium charge $\bar Z\sub{VSG}$ such that it collides with
ions as frequently as it collides with electrons. That is, $\bar
Z\sub{VSG}$ is such that the rate coefficients (units of cm$^3$
s$^{-1}$) $\alpha_{\rm VSG,ion}$ and $\alpha_{\rm VSG,e}$---as given by
equations 9 and 10 of PBC11---are equal (see Figures 3 and 4 of PBC11).
Under this approximation, 
the rate at which an ion (CII or SII) neutralizes by colliding with VSGs is
simply $n\sub{VSG} \alpha\sub{VSG}$, where $\alpha\sub{VSG} =
\alpha\sub{VSG,ion} = \alpha\sub{VSG,e}$.  
The values of $\alpha\sub{VSG}$ so derived differ by 50--70\%
depending on whether the dominant ion is CII (atomic weight 12) or SII (atomic weight 32); for simplicity we take a mean value.
Given our choices for $T$
and $s$, and assumed sticking coefficients between VSGs and electrons/ions of 1, we find that for our standard model
$\bar Z\sub{VSG} \approx -1$ in units of the electron
charge,\footnote{Our $\bar Z$ differs slightly from the
  true average $\langle Z \rangle$ calculated by PBC11.}  and the
corresponding $\alpha\sub{VSG} \approx 2.5\e{-6}$ cm$^3$/s.

A few parting comments about our treatment of VSGs: first, in deciding
how VSGs attenuate FUV radiation, we do not make any explicit
assumption about the grain size distribution. Our proxy for flux
attenuation $\Nvsg$ is the column at which FUV radiation is absorbed
by grains of all sizes.  Flux attenuation by VSGs will play a
significant role in determining the extent of MRI activity (Section
\ref{sec_sensitivity}).  Second, in deciding how ions recombine on
VSGs, we do make an explicit assumption about grain size, namely we
take $s\sub{VSG} = 0.01 \mu$m.  Ignoring larger grain sizes maximizes
the number abundance and geometric surface area of VSGs, and thus
maximizes their relevance for charge balance. With this choice for grain size,
ion recombination on VSGs is competitive with other recombination
pathways for our standard model ($\Nvsg = 10^{22}$ cm$^{-2}$), but is
negligible for more dust-depleted models ($\Nvsg > 10^{22}$
cm$^{-2}$).

\subsection{Polycyclic Aromatic Hydrocarbons (PAHs)}
Ion recombination on PAHs is treated analogously to ion recombination
on VSGs. Each PAH has radius 6 $\AA$; an electron (ion)
sticking coefficient of 0.1 (1); and thus an average charge $\bar
Z\sub{PAH} \approx -0.05$ at $a = 3$ AU.  The corresponding ion-PAH rate
coefficient $\alpha\sub{PAH} \approx 1\e{-8}$ cm$^3$/s.

For our standard (fiducial) model, we take the number abundance of
PAHs relative to hydrogen nuclei to be $x_{\rm{PAH}}\equiv n_{\rm
  PAH}/n\sub{tot} = 10^{-8}$. This is their maximum plausible
abundance, as inferred by PBC11.  For this PAH abundance, the ion
recombination rate on PAHs will turn out to be only comparable to the
ion recombination rate with free electrons in FUV-ionized layers (the
contribution from PAHs was much more significant for the deeper and
more poorly ionized layers considered in the X-ray model of
PBC11). Thus we expect the thickness of the FUV-irradiated, MRI-active
layer to be insensitive to PAHs, a finding we highlight at the end of Section
\ref{sec_sensitivity}.

Note that under our simplified treatment of recombination
on condensates, the net charge of our system is
not zero: VSGs and PAHs carry a net negative charge, which
together with the contribution from free electrons is not
balanced by positive CII and SII ions.  However the deviation from charge
neutrality (which was not present in PBC11 because that study
accounted explicitly for collisional charging of condensates by ions
and electrons) is small. That is, we have verified that $\bar Z\sub{PAH}n_{\rm
  PAH} + \bar Z\sub{VSG}n_{\rm VSG}$ is much less than the electron
number density over the domain of our calculation.

\subsection{H$_2$: Photodissociation and Re-Formation on Grains}
Molecular hydrogen is photodissociated by FUV photons in two steps: a
photon of wavelength $912 \AA < \lambda < 1109 \AA$ sends the molecule into the first-excited electronic state, and then in the subsequent radiative
decay, there is a $p=23$\% probability that the molecule lands in a
vibrationally excited state where it becomes unbound
\citep{Dalgarno:1970p8972}.  Our calculation of the FUV dissociation
rate of H$_2$ molecules follows that of \citet{deJong:1980p8894},
according to which there are $\eta\sub{L}=60$ vibrationally split line
transitions connecting the ground and first-excited electronic
states. These lines constitute the so-called Lyman band.

The dissociation rate is
\begin{equation}\label{Rphoto}
	R_{\rm diss}=p n\sub{H_2} \eta\sub{L} \frac{\pi e^2 f / (m\sub{e}c)}{\nu} 
\frac{(2/3)L\sub{FUV}}{4\pi a^2 h \nu} \beta\sub{VSG} \beta\sub{H_2} \,.
\end{equation}
All lines of frequency $\nu$ (energy $h\nu \approx 12$ eV) in the
Lyman band are assumed to be equally spaced and to have equal
oscillator strengths $f=4.6\e{-3}$. The factor $\pi e^2 f / (m\sub{e}
c)$ is the frequency-integrated cross section of a single line, where
$e$ and $m\sub{e}$ are the charge and mass of an electron, and $c$ is
the speed of light.  Thus
$\eta\sub{L} [\pi e^2 f / (m\sub{e} c)]/\nu
\equiv \sigma\sub{eff}$
approximates the effective cross section of
the entire Lyman band, averaged over the incident broadband spectrum
of frequency width $\sim$$\nu$. The factor of $2/3$ accounts
approximately for the fraction of $L_{\rm FUV}$ (as we have defined it
for $912 \AA < \lambda < 1198 \AA$) that covers the Lyman band (see
Section \ref{fuv}).

The factor $\beta\sub{VSG} = \exp [-N_{\rm tot}/(\theta \Nvsg)]$ accounts for
attenuation of FUV radiation by VSGs (recall that $N_{\rm tot}$
is the vertical column so that $N_{\rm tot}/\theta$ is the column
parallel to the incident beam of radiation). The factor $\beta\sub{H_2}$
accounts for attenuation of H$_2$-dissociating photons by intervening H$_2$
\citep{deJong:1980p8894}:
\begin{equation}
\beta\sub{H_2}=\left(\frac{1}{\tau\sub{H_2} \left[\ln \left(\tau\sub{H_2}/ \sqrt{\pi} \right)   \right]^{1/2}} + \sqrt{\frac{b}{\tau\sub{H_2}}}\right) \mathrm{erfc}\left(\sqrt{\frac{\tau\sub{H_2} b}{\pi \Delta^2}}   \right) \label{beta_SS}
\end{equation}
where $\mathrm{erfc}$ is the complementary error function. 
The core of each line is assumed to be Doppler broadened
with frequency width $\Delta\nu\sub{D} \equiv c\sub{s} \nu / c$;
then
\begin{equation}
\tau\sub{H_2} = \frac{\pi e^2 f}{m\sub{e}c \Delta\nu\sub{D}} N\sub{H_2}/\theta
\end{equation}
is a measure of the optical depth at line center along the incident beam.
The Voigt parameter
\begin{equation}
	b  = \gamma/(4 \pi \Delta \nu\sub{D})%
\end{equation}
compares the natural line width $\gamma = 1.16\e{9}$ s$^{-1}$
to the Doppler width, and the non-dimensional parameter
\begin{equation}
	\Delta = \Delta \nu\sub{L} / (2 \eta\sub{L} \Delta \nu\sub{D})%
\end{equation}
compares the inter-line spacing to the Doppler width, where $\Delta \nu\sub{L}=6\e{14}$ Hz is the frequency width of the entire Lyman band.  Equation
(\ref{beta_SS}) applies only for $\tau\sub{H_2} \gg 1$, a condition
which is valid over the domain of our calculation.

Although the photodissociating continuum for H$_2$
and the photoionizing continuum for CI nearly overlap
in the FUV, we can safely neglect shielding of H$_2$ by CI because in our model the H$_2$ dissociation front occurs at a higher
altitude (lower $N\sub{tot}$) than the C ionization front. 
By the same token, shielding of CI by H$_2$ cannot be neglected
(see Section \ref{carbon} where it is accounted for
by the factor $\beta\sub{C,H_2}$).

Molecular hydrogen forms when two H atoms combine on a grain surface,
releasing the heat of formation to the grain lattice
\citep{Gould:1963p9112}.  The rate of formation is given by
\begin{equation}\label{Rform}
	R_{\rm form}=\frac{1}{2} n\sub{H} n\sub{VSG} \pi s\sub{VSG}^2 v\sub{H} \eta %
\end{equation}
where $n_{\rm H}$ and $v\sub{H}$ are the number density and mean
thermal speed of atomic hydrogen, respectively, and $\eta \sim 0.2$ is
the formation efficiency on olivine grains \citep{Cazaux:2010p8085}.

At each depth in our 1D model, we solve for $n_{\rm H}/n_{\rm H_2}$
using the equilibrium condition (\ref{quality_diss}):
\begin{equation}
R_{\rm diss} = R_{\rm form} \,. \label{quantity_diss}
\end{equation}

\subsection{Carbon: Abundance and Ionization Equilibrium}\label{carbon}

We take the abundance of atomic carbon in the gas phase to be
$x\sub{C} \equiv n\sub{C}/n\sub{tot} = 10^{-4} \epsilon$, where the
dimensionless factor $\epsilon$ accounts for the sequestering of
carbon into grains. For our standard model, $\epsilon = 1$, which gives a
gas phase carbon abundance similar to that of the diffuse
interstellar medium \citep{Jenkins:2009p6986}.  By using [CII] 158 $\mu$m line fluxes
from disks as measured by the \textit{Herschel} satellite
\citep{Pinte:2010p8649}, we estimate a lower bound on $\epsilon$ of
$\sim$$1/30$.

Equation (\ref{quality_carbon}) for photoionization equilibrium of
carbon is quantified as
\begin{eqnarray}\label{quantity_c}
	\frac{(2/3)L_{\rm{FUV}}}{4 \pi a^2 h \nu} n_{\rm{CI}} \sigma_{\rm{bf,C}} \beta_{\rm{C}} \beta\sub{C,H_2} \beta\sub{VSG} \nonumber\\
	=n_{\rm{CII}} \left(n_{\rm{e}}\alpha_{\rm{rec,C}}+n_{\rm{PAH}}\alpha_{\rm{PAH}}+n_{\rm{VSG}}\alpha_{\rm{VSG}}\right), 
\end{eqnarray}
where $h\nu \approx 12$ eV is the typical FUV photon energy,
$n_{\rm{CI}} \, (n_{\rm{CII}} = n_{\rm C}-n_{\rm CI})$ is the neutral
(ionized) carbon number density, $n_{\rm{e}}$ is the
electron number density, $\sigma_{\rm{bf,C}} = 10^{-17}$ cm$^2$ is the
bound-free cross section for CI \citep{Cruddace:1974p7788}, and
$\alpha\sub{rec,C} = 5.3\e{-12} (T/240\,\rm{K})^{-0.6}$ cm$^3$/s is the
radiative recombination rate coefficient for CII
\citep{Pequignot:1991p9118}.  
The factor of $2/3$ accounts for the fraction of $L_{\rm FUV}$ (as we have
defined it in Section \ref{fuv}) covered by the photoionizing continuum for
CI. The factor $\beta\sub{C}= \exp (-N_{\rm
  CI}\sigma\sub{bf,C}/\theta)$ accounts for self-shielding of carbon,
where $N_{\rm CI}$ is the vertical column of CI. The factor
$\beta\sub{C,H_2}$ accounts for shielding of carbon by the Lyman band
transitions of molecular hydrogen \citep{deJong:1980p8894}:
\begin{equation}
\beta\sub{C,H_2}=\left(1+\frac{\tau\sub{H_2}b}{\pi \Delta^2}\right)^{-1} \mathrm{exp}\left(-\frac{\tau\sub{H_2}b}{\pi \Delta^2}\right).	
\end{equation}
This factor drops significantly when the H$_2$ column
$N\sub{H_2}/\theta \gtrsim 10^{22}$ cm$^{-2}$---large enough for the
Lorentzian wings of each Lyman line to blacken the entire Lyman band.
As noted previously, all the FUV photons that can ionize CI
($1104 \AA > \lambda > 912 \AA$) can also dissociate H$_2$
($1109 \AA > \lambda > 912 \AA$).
This is true for C but not for other elements like S.

We neglect shielding of CI by SI because the S ionization front lies
at a greater depth than that of the C ionization front (because S is 
a factor of 10 less abundant than C everywhere).

\subsection{Sulfur: Abundance and Ionization Equilibrium}\label{sulfur}

The abundance of gas-phase atomic sulfur is $x\sub{S} \equiv
n\sub{S}/n\sub{tot} = 10^{-5} \epsilon$.  For our standard model,
$\epsilon = 1$, which yields a sulfur abundance similar to that of the
diffuse interstellar medium (Jenkins 2009).  For simplicity, we assume
that the depletion factor $\epsilon$ for sulfur is the same as that
for carbon, so that $x\sub{S} / x\sub{C} = 0.1$ for all our
models. The lower abundance of sulfur relative to carbon implies that
S-ionizing radiation will penetrate to greater depths than C-ionizing
radiation (see, e.g., the order-of-magnitude estimate in PBC11),
implying a deeper and more massive MRI-active layer.

Equation (\ref{quality_sulfur}) for photoionization equilibrium of
sulfur is quantified as
\begin{eqnarray}\label{quantity_s}
	\frac{(1/3)L_{\rm{FUV}}}{4 \pi a^2 h \nu} n_{\rm{SI}} \sigma_{\rm{bf,S}} \beta_{\rm{S}} \beta\sub{VSG} \nonumber\\
	=n_{\rm{SII}} \left(n_{\rm{e}}\alpha_{\rm{rec,S}}+n_{\rm{PAH}}\alpha_{\rm{PAH}}+n_{\rm{VSG}}\alpha_{\rm{VSG}}\right), 
\end{eqnarray}
where $h\nu \approx 11$ eV,
$\sigma_{\rm bf,S} \approx 5.5 \times 10^{-17}$ cm$^2$
is the photoionizing cross-section for S
\citep{Cruddace:1974p7788},
$\alpha\sub{rec,S} = 4.9\e{-12} (T/240\,\rm{K})^{-0.63}$ cm$^3$/s
is the radiative recombination rate
coefficient for SII \citep{Aldrovandi:1973p10605,Gould:1978p10616},
and $n_{\rm{SI}} \, (n_{\rm{SII}} = n_{\rm
  S}-n_{\rm SI})$ is the neutral (ionized) sulfur number density. 
The factor $\beta\sub{S}= \exp (-N_{\rm SI}\sigma\sub{bf,S}/\theta)$
accounts for self-shielding of sulfur, where $N_{\rm SI}$ is the
vertical column of SI. 

The factor of 1/3 accounts for the fraction of $L_{\rm FUV}$ (as we have
defined it for $912 \AA < \lambda < 1198 \AA$ in Section \ref{fuv})
that can ionize S but not be absorbed by CI and H$_2$.
Technically there should be a second term on the left-hand
side of (\ref{quantity_s}) that is proportional to
$(2/3)L_{\rm FUV}$, to account for radiation that can ionize
S and be absorbed by CI and H$_2$. But this term can be dropped
because it does not significantly influence the location of the S ionization
front; the term is negligibly small there, at a greater depth
than the C ionization front.
Note that the governing equation for sulfur
is coupled to the other equations for C and H$_2$ only via the
electron density $n_{\rm e} = n_{\rm CII} + n_{\rm SII}$.

\subsection{Numerical Method of Solution}

Our calculation starts at $N\sub{tot}=10^{18}$ cm$^{-2}$, under the
assumption that the incident flux $L\sub{FUV}/4\pi a^2$ is not
attenuated at this column. We take $10^3$ logarithmically spaced steps
to $N\sub{tot} = 10^{24}$ cm$^{-2}$. At each step, we solve for the
ratio $n\sub{H}/n\sub{H_2}$ using the balance condition
(\ref{quantity_diss}). The self-shielding factor $\beta\sub{H_2}$
depends on $N_{\rm H_2}$ and so we keep a running integral of $n_{\rm
  H_2}$, i.e., $N_{\rm H_2} = \sum_i (n_{\rm H_2}/n\sub{tot})_i
(\Delta N\sub{tot})_i$.

By approximating $n\sub{e}=$ max$(n\sub{CII},n\sub{SII})$, we may
solve analytically for the ionization states of C and S. At low
$N\sub{tot}$, nearly all C and S are ionized and $n\sub{e} =
n\sub{CII}$ with an error of $x\sub{C}/x\sub{S}\approx 10$\%.  Equation
(\ref{quantity_c}) becomes a quadratic for $n_{\rm CII}/n_{\rm CI}$,
and $n\sub{SII}/n\sub{SI}$ is determined by substituting $n\sub{CII}$
for $n\sub{e}$ in equation (\ref{quantity_s}).

At $N\sub{tot}$ past the C ionization front, SII becomes the dominant
ion and we set $n\sub{e} = n\sub{SII}$. Now equation (\ref{quantity_s})
becomes a quadratic for $n\sub{SII}/n\sub{SI}$.
The resultant value for $n\sub{SII}$ is substituted for $n\sub{e}$ in
equation (\ref{quantity_c}) to solve for $n\sub{CII}/n\sub{CI}$.

At each $N\sub{tot}$, we keep track of the running columns $N_{\rm CI}
= \sum_i (n_{\rm CI}/n\sub{tot})_i (\Delta N\sub{tot})_i$ and $N_{\rm
  SI} = \sum_i (n_{\rm SI}/n\sub{tot})_i (\Delta N\sub{tot})_i$ to
compute the self-shielding factors $\beta\sub{C}$ and $\beta\sub{S}$,
respectively.

\section{RESULTS FOR MRI-ACTIVE SURFACE DENSITIES} \label{results}

In Section \ref{fronts} we describe how the H$_2$:H, CI:CII, and
SI:SII ratios vary with depth. In Section \ref{AmRe} we estimate the
surface density $\Sigma^{*}$ of the layer that can be MRI-active. In
Section \ref{sec_sensitivity} we perform a parameter study of how
$\Sigma^{*}$ varies with system properties. The MRI-active surface
densities $\Sigma^\ast$ reported in Section \ref{sec_sensitivity} are
those resulting from FUV ionization only. 
In Section \ref{sec_hall} we
check that the Hall effect is not significant for FUV-ionized surface
layers. All our calculations neglect turbulent mixing of plasma from
disk surface layers into the disk interior. In Section \ref{sec_trec},
we try to assess whether this neglect is justified, by comparing the
dynamical timescale with the timescale over which the FUV-irradiated
layer equilibrates chemically.

\begin{figure} %
\epsscale{0.90}
\plotone{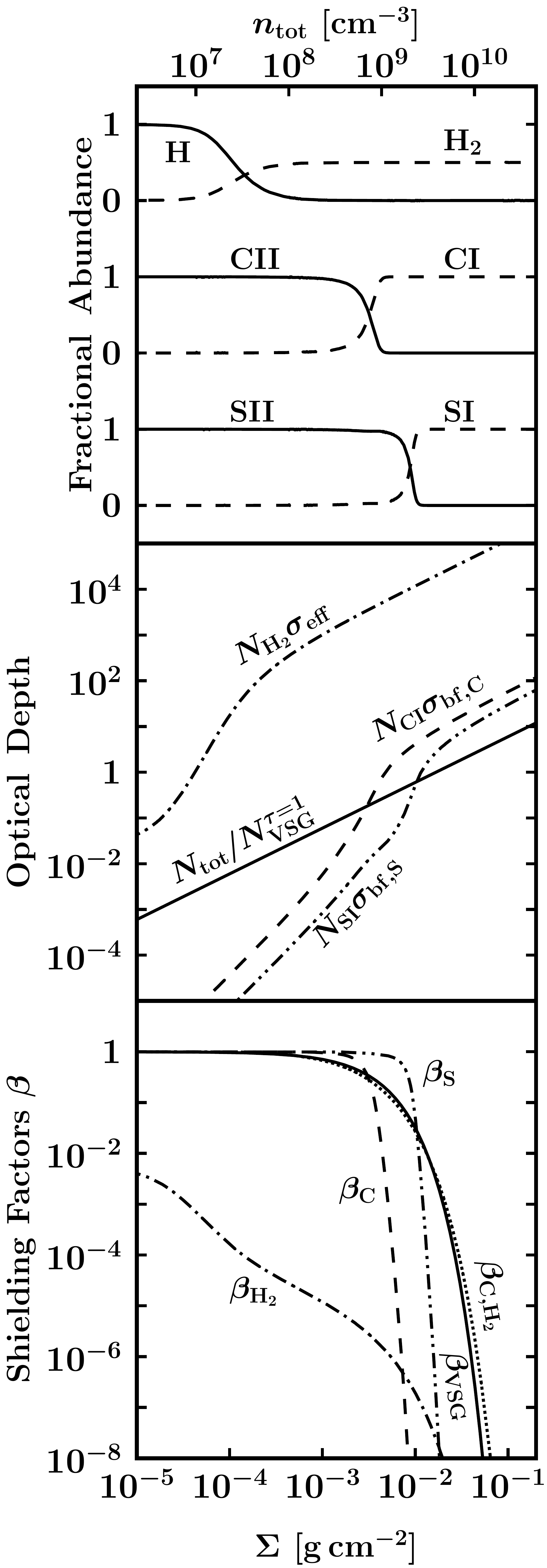}
\caption{H$_2$ photodissociation front, and C and S ionization
    fronts, for our standard model at $a=3$ AU.  Results are given as
    functions of the vertical surface density $\Sigma$, measured
    perpendicular to the disk midplane (the column actually penetrated
    by radiation parallel to the incident beam direction is
    $\Sigma/\theta$).  The corresponding number density of H nuclei is
    given on the top x-axis. {\it Upper panel}: Normalized abundances
    of atomic H, H$_2$, CII, CI, SII, and SI. The H$_2$
    photodissociation front occurs at the highest altitude above the
    midplane, and the S ionization front occurs at the lowest. {\it
      Middle panel}: Optical depths to FUV radiation presented by
    H$_2$ (averaged over the Lyman band), CI, SI, and dust
    (VSGs). {\it Lower panel}: Shielding factors by which FUV
    radiation is attenuated: self-shielding of H$_2$
    ($\beta\sub{H_2}$); self-shielding of C ($\beta\sub{C}$);
    shielding of C by H$_2$ ($\beta\sub{C,H_2}$); self-shielding of S
    ($\beta\sub{S}$); and shielding by very small grains
    ($\beta\sub{VSG}$). }
\label{fig_fronts}
\end{figure}

\subsection{Photodissociation and Ionization Fronts} \label{fronts}

In the top panel of Figure \ref{fig_fronts}, we show the relative
abundances of species as a function of vertical mass column 
$\Sigma$. High in the atmosphere, hydrogen is mostly in atomic form
and nearly all of the carbon and sulfur are ionized.
Deeper down, starting at $\Sigma\sim 3\e{-5}$ g cm$^{-2}$, atomic H
gives way to H$_2$. 
At the H$_2$:H front the Doppler broadened cores of the various H$_2$
Lyman transitions have become optically thick. Far-UV photons in the
wings of the Lyman transitions can stream past the H$_2$:H front and
continue to ionize C and S at greater depths.

The CI:CII ionization front is located at $\Sigma\sim 3\e{-3}$ g
cm$^{-2}$, where the optical depth to carbon ionizing photons is of
order unity. At this front, both shielding of carbon by H$_2$
(quantified by $\beta\sub{C,H_2}$) and carbon self-shielding
($\beta\sub{C}$) are becoming significant; see the bottom panel of
Figure \ref{fig_fronts}.

Far-UV photons having $1198 \AA > \lambda > 1109 \AA$ interact with
neither H$_2$ nor C, but can ionize S. These photons penetrate both
H$_2$ and C fronts, and ionize S at greater depths. The SI:SII
ionization front is located at $\Sigma\sim 1\e{-2}$ g cm$^{-2}$.  In
this fiducial model, the location of the S ionization front is
determined more by flux attenuation by very small grains
than by self-shielding by sulfur;
at the front, $\beta\sub{VSG} \sim 0.1$ and $\beta\sub{S} \sim 1$.
Lower but still observationally realistic grain abundances are
considered in Section \ref{sec_sensitivity}, where we will
find that the S ionization front can be as deep as
$\Sigma \sim 1\e{-1}$ g cm$^{-2}$.

\begin{figure} %
\epsscale{1.0}
\plotone{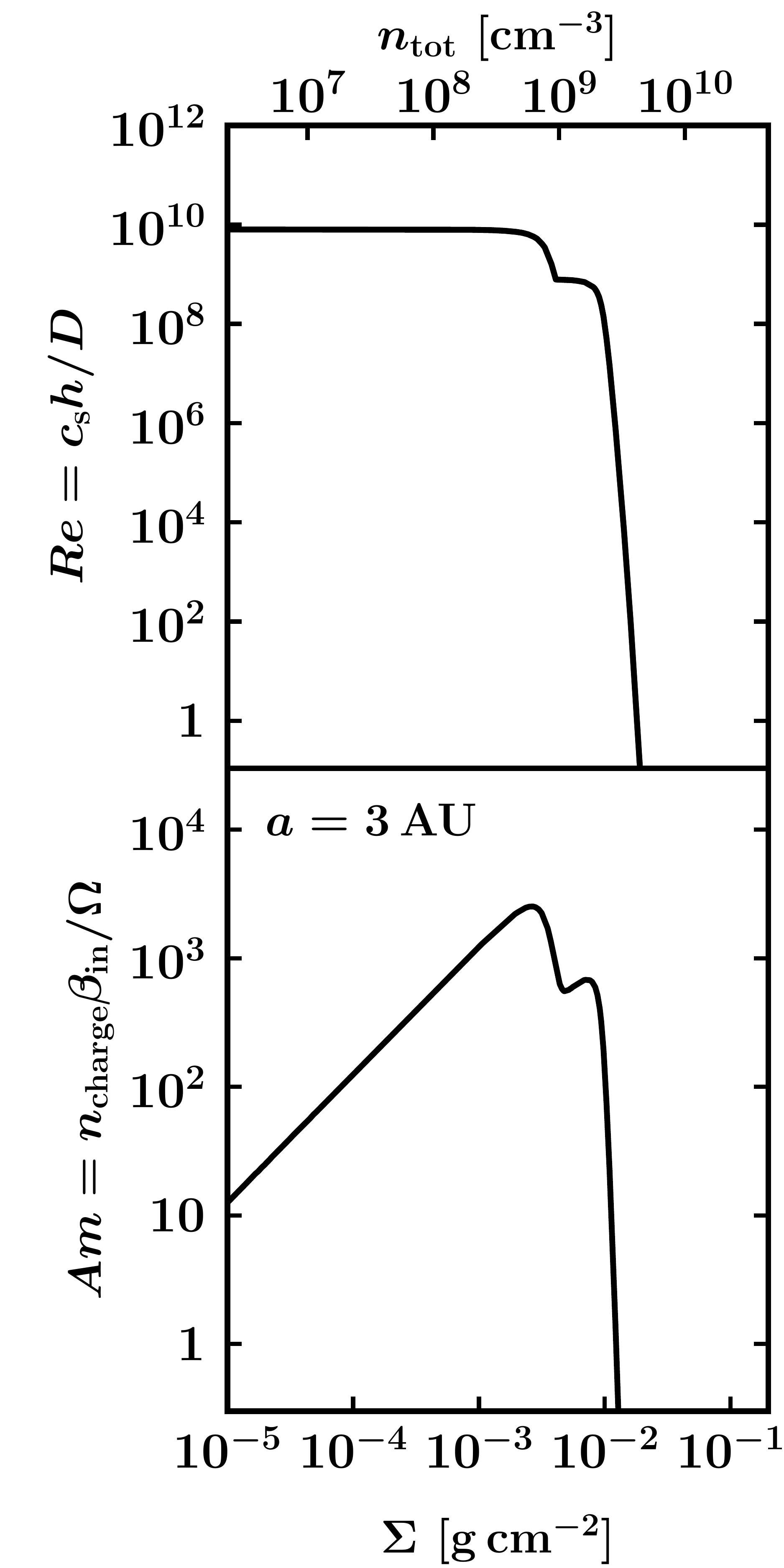}
\caption{Magnetic Reynolds number $Re$ and ambipolar diffusion number
  $Am$ versus vertical surface density $\Sigma$ for standard model
  parameters and $a=3$ AU.} 
\label{fig_AmRe3AU}
\end{figure}

\begin{figure} %
\epsscale{1.0}
\plotone{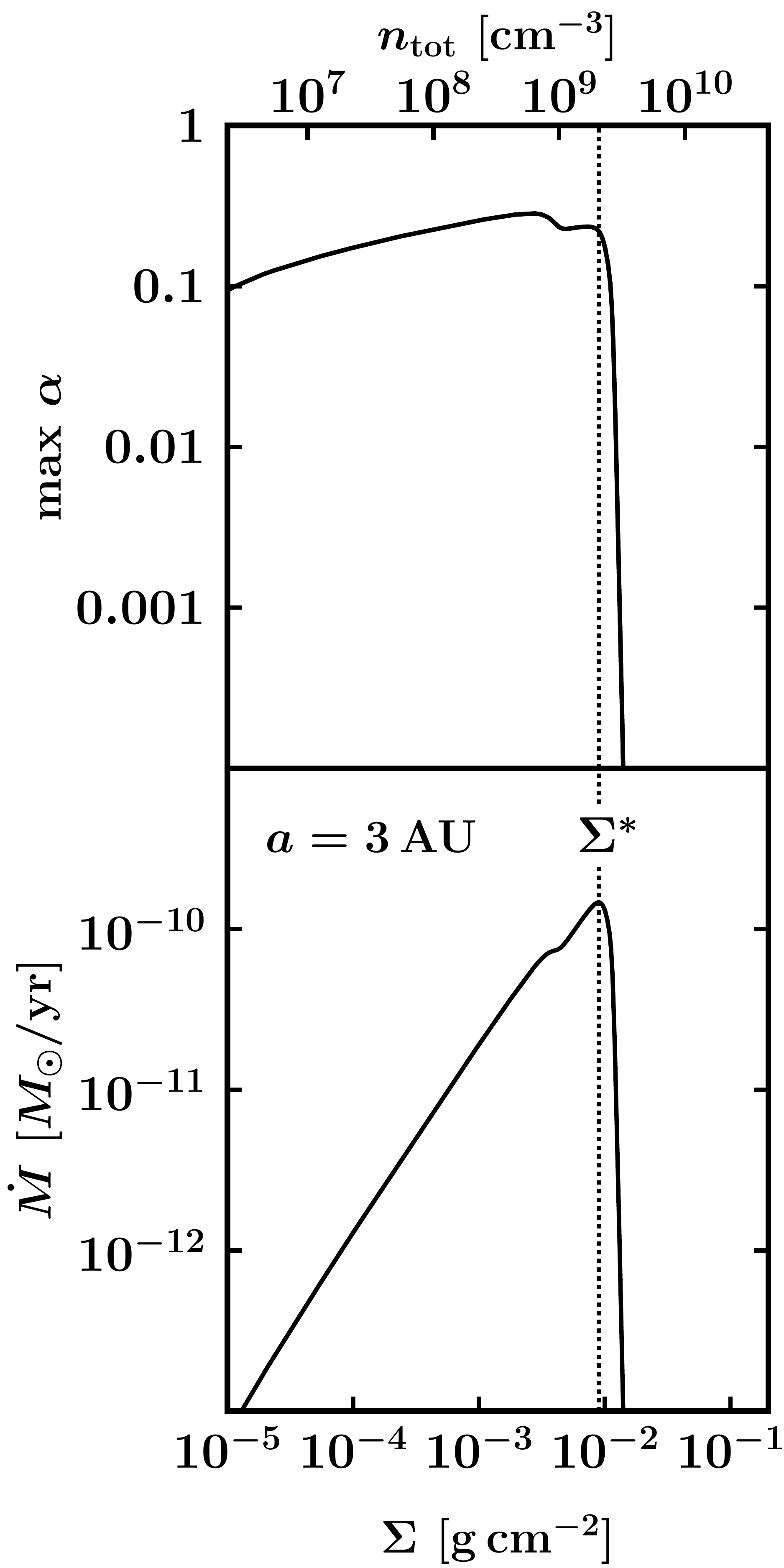}
\caption{Maximum transport parameter $\alpha$ and mass
    accretion rate $\dot{M}$ versus vertical surface density $\Sigma$
    for standard model parameters and $a=3$ AU. The value of $\max \alpha$
    is computed using the empirical relation (\ref{maxalpha}) of BS11.
    A vertical dotted line
    marks the MRI-active surface density $\Sigma^{*}$ at which, by definition,
    $\dot{M}$ peaks. The value of $\Sigma^\ast$ in FUV-ionized layers is
    essentially set by the SI:SII ionization front. It can be as large
    as $\sim$0.1 g/cm$^2$ for parameter choices other than those
    assumed here; see Figure \ref{fig_sensitivity}.}
\label{fig_AlphaMdot}
\end{figure}

\vspace{0.3in}

\subsection{MRI-Active Surface Density $\Sigma^{*}$}\label{AmRe}

We measure the extent of the MRI-active column by means of the
magnetic Reynolds number $Re$ and the ion-neutral collision rate
$Am$. Figure \ref{fig_AmRe3AU} shows both dimensionless numbers as a
function of $\Sigma$ for our standard model at $a=3$ AU.

Above the CI:CII ionization front, at $\Sigma \lesssim 3\e{-3}$ g
cm$^{-2}$, the value of $Re$ remains constant at $\sim$$10^{10}$
because $x\sub{e}$ saturates at $10^{-4}$ where nearly all the carbon
is singly ionized. In this region, $Am \propto \Sigma$ because $Am
\propto n\sub{tot}$. The value of $Am$ peaks at $\sim$$3\times 10^3$
near the C ionization front.
Both $Am$ and $Re$ fall just past the C ionization
front, but not by more than an order of magnitude: S, which is
10$\times$ less abundant than C, remains fully ionized out to the S
ionization front at $\Sigma \sim 10^{-2}$ g/cm$^2$. Past the S
ionization front, virtually all FUV photons are used up; the charge
density, and by extension both dimensionless numbers, drop sharply.

We define the MRI-active surface density $\Sigma^\ast$ as follows.
Bai \& Stone (2011) find from an extensive series
of numerical simulations that for a given $Am$,
the maximum value of $\alpha$ is given by 
\begin{equation} \label{maxalpha}
\max \alpha = \frac{1}{2} \left [ \left ( \frac{50}{Am^{1.2}} \right )^2 +\left ( \frac{8}{Am^{0.3}}+1 \right )^2 \right ]^{-1/2} \,.
\end{equation}
From the value of $\max \alpha$ so computed, we evaluate the mass accretion rate
\begin{equation} \label{mdot_define}
\dot{M} = 2 \times 3 \pi \Sigma \nu = 6 \pi \Sigma \times \max \alpha \times \frac{kT}{\mu \Omega}
\end{equation}
as a function of $\Sigma$, where $\nu = \alpha c\sub{s} h$ is the
viscosity, $\mu = 4 \times 10^{-24}$ g is the mean molecular weight,
$k$ is Boltzmann's constant, 
and the prefactor of 2 accounts for the top and bottom
surfaces of the disk. Note that $\max \alpha$ is a function of $Am$,
which in turn is a function of $\Sigma$ that we have computed (Figure
\ref{fig_AmRe3AU}).  We define $\Sigma^\ast$ as that value of $\Sigma$
for which $\dot{M}$ peaks; see Figure \ref{fig_AlphaMdot}, which plots equation (\ref{mdot_define}). According
to this definition, $\Sigma^{*} \approx 1\e{-2}$ g/cm$^{2}$. At this column,
according to Figure \ref{fig_AmRe3AU},
$Re$ remains larger than $Re^*$, demonstrating that ambipolar diffusion
limits MRI activity more than Ohmic dissipation does, although the effect is slight (the dominance of
ambipolar diffusion over Ohmic dissipation was much more pronounced
for the X-ray ionized layers analyzed by PBC11). Referring
back to Figure \ref{fig_fronts}, we see that $\Sigma^\ast$ corresponds
essentially to the SI:SII ionization front.

 \begin{figure*} %
\epsscale{1.0}
\plotone{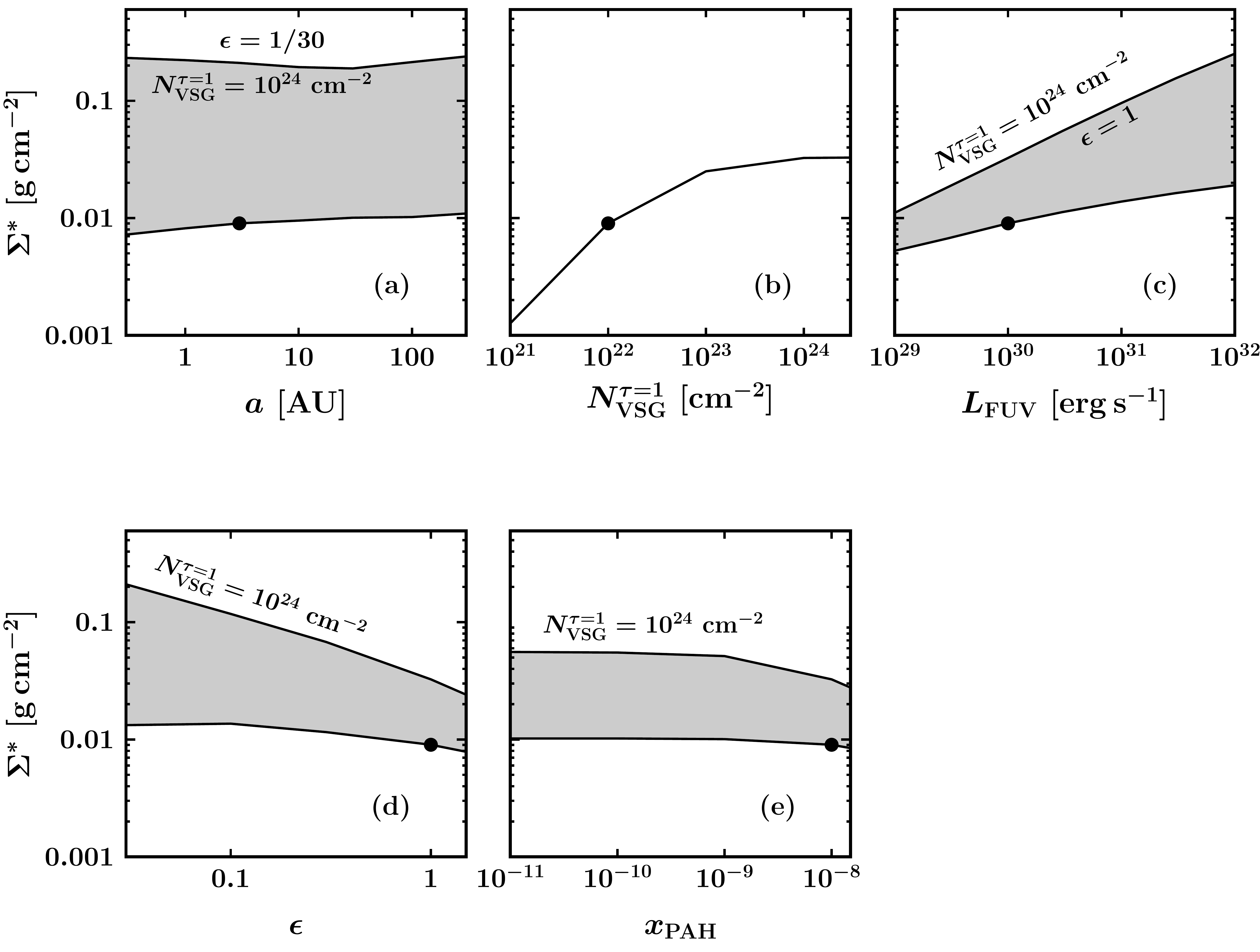}
\caption{How $\Sigma^{*}$ varies with (a) stellocentric
  distance $a$; (b) our proxy $\Nvsg$ for the abundance of very small
  grains (see equation \ref{vsg_abun}); (c) FUV luminosity $L\sub{FUV}$; (d) the gas-phase abundance of carbon and sulfur $\epsilon$
  (normalized so that $\epsilon = 1$ corresponds to near-solar
  abundances); and (e) the abundance of charge-absorbing PAHs.  For a given solid curve in a given panel, only the one
  parameter on the x-axis is varied while all other parameters, unless
  otherwise annotated on the figure, are held fixed at their standard
  values.  The upper bounding curves in panels (a), (c), (d), and (e) are
  calculated assuming $\Nvsg = 10^{24}$ cm$^{-2}$, the lowest grain
  abundance considered.  In panel (a), the upper bounding curve
  assumes $\epsilon = 1/30$, which we estimate to be the lowest
  possible value still consistent with observations of far-infrared
  [CII] emission from disks
  (\citealt{Pinte:2010p8649,Sturm:2010p8660}). A dot in each panel
  marks the standard (fiducial) model.}
\label{fig_sensitivity}
\end{figure*}

\subsection{Parameter Survey, Including Sensitivity to PAHs}\label{sec_sensitivity}

In Figure \ref{fig_sensitivity} we explore the sensitivity of
$\Sigma^{*}$ to various parameters. For a given solid curve in a given
panel, only the one parameter on the x-axis is varied while all others
are held fixed. Labels on a curve indicate the values to which certain
parameters are fixed when computing that curve. If a parameter is not
labeled, its value was set to the standard value.

Over most of parameter space, the MRI-active surface density
$\Sigma^\ast$ in FUV-ionized layers is of order $10^{-2}$--$10^{-1}$
g/cm$^{2}$.  The MRI-active surface density is set by the depth of the
sulfur ionization front.  Very small grains (VSGs) reduce the depth of
this front by absorbing FUV radiation (and also by enhancing the ion
recombination rate, although this is not a dominant effect for
FUV-ionized layers where the electron density and thus the radiative
recombination rate are high).  At their standard abundance---which is
probably near their maximum abundance (Section \ref{vsg_param})---VSGs
shield sulfur from ionizing radiation more than sulfur shields
itself; compare $\beta_{\rm S}$ and $\beta_{\rm VSG}$ in Figure
\ref{fig_fronts}.

As $\Nvsg$ increases from $10^{22}$ to $10^{24}$ cm$^{-2}$ (i.e., as
the grain abundance decreases), $\Sigma^\ast$ increases by a factor of
3 (Figure \ref{fig_sensitivity}b). Now most of the shielding of sulfur
from FUV radiation is provided by sulfur itself, not by VSGs. When
grains are insignificant ($\Nvsg = 10^{24}$ cm$^{-2}$), $\Sigma^\ast$
depends on $L_{\rm FUV}$ and the carbon/sulfur abundance $\epsilon$
more strongly than in the case when grains are significant absorbers
of FUV radiation (Figure \ref{fig_sensitivity}c,
\ref{fig_sensitivity}d).  The dependences in the case of low grain
abundance resemble those calculated by PBC11 in their simple
FUV-Str\"omgren model: $\Sigma^\ast \propto L_{\rm FUV}^{1/2}$ and
$\Sigma^\ast \propto 1/\epsilon$.

The MRI-active thickness $\Sigma^\ast$ in FUV-ionized layers
hardly varies with disk radius $a$ (Figure \ref{fig_sensitivity}a),
because the decrease in ionizing flux with increasing $a$ is
compensated by the lengthening dynamical time $\Omega^{-1}$ (to which
$Am$, the factor determining $\Sigma^\ast$, is proportional).

Finally, $\Sigma^\ast$ is not sensitive to the abundance of PAHs
(Figure \ref{fig_sensitivity}e).  For the large electron fractions
$x_{\rm e} \approx 10^{-5}$--$10^{-4}$ generated by FUV ionization of
carbon and sulfur, the primary channel for charge neutralization is
radiative recombination with free electrons. Only at the
highest PAH abundances does the ion recombination rate with PAHs
($x\sub{PAH} \alpha\sub{PAH}$) approach the ion recombination rate
with electrons ($x\sub{e} \alpha\sub{rec,S}$).  By contrast, X-ray
ionized gas has much lower electron fractions
and is much more susceptible to charge recombination on
small condensates like PAHs (PBC11).

\begin{figure} %
\epsscale{1.0}
\plotone{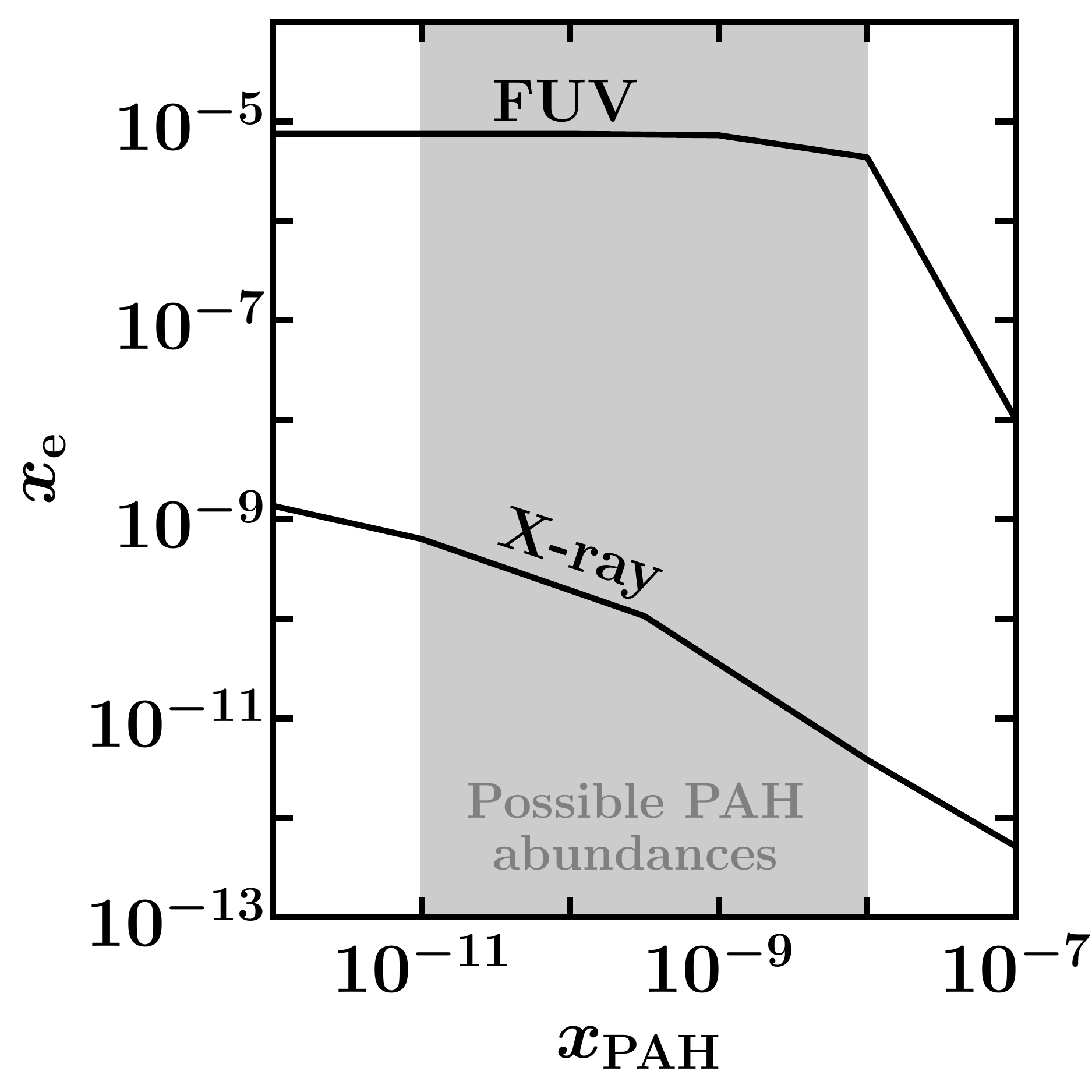}
\caption{Electron fraction as a function of PAH abundance for FUV and
  X-ray ionized surface layers. The FUV curve is computed for our
  standard model parameters ($a=3$ AU, $L\sub{FUV}= 10^{30}$ erg/s,
  $\Nvsg=10^{22}$ cm$^{-2}$, $\epsilon=1$) at $\Sigma \approx 0.01$ g
  cm$^{-2}$. The large abundance of electrons generated in FUV-ionized
  layers is immune to the effects of charge recombination on PAHs,
  over the range of plausible PAH abundances (shaded in grey; PBC11).
  We contrast this behavior with the X-ray curve taken from Figure 7
  of PBC11 ($a=3$ AU, $L\sub{X} = 10^{29}$ erg/s, metal abundance
  $x\sub{M}=10^{-8}$, $\Sigma = 0.3$ g cm$^{-2}$), which shows that
  PAHs can reduce electron fractions in X-ray-irradiated layers by two
  orders of magnitude.}
\label{fig_xe_vs_xPAH}
\end{figure}

The last point is echoed in Figure \ref{fig_xe_vs_xPAH}, which shows
how the free electron fraction varies with PAH abundance in both
FUV-ionized and X-ray-ionized surface layers.  Each curve is computed
at a fixed surface density characteristic of each layer: $\Sigma
\approx 0.01$ g/cm$^2$ for the FUV-irradiated layer, and $\Sigma
\approx 0.3$ g/cm$^2$ for the X-ray-irradiated layer. For the
FUV-ionized layer, $x\sub{e}$ remains constant over the range of
plausible PAH abundances $10^{-11} \lesssim x\sub{PAH} \lesssim
10^{-8}$. For the X-ray-ionized layer, $x\sub{e}$ varies by about two
orders of magnitude over the same range of PAH abundance; $x\sub{e}
\propto 1/x\sub{PAH}$ for $x\sub{PAH} \gtrsim 5 \times 10^{-10}$ (see
Section 3.2.1 of PBC11).

\subsection{Hall Diffusion} \label{sec_hall}

\begin{figure} %
\epsscale{1.0}
\plotone{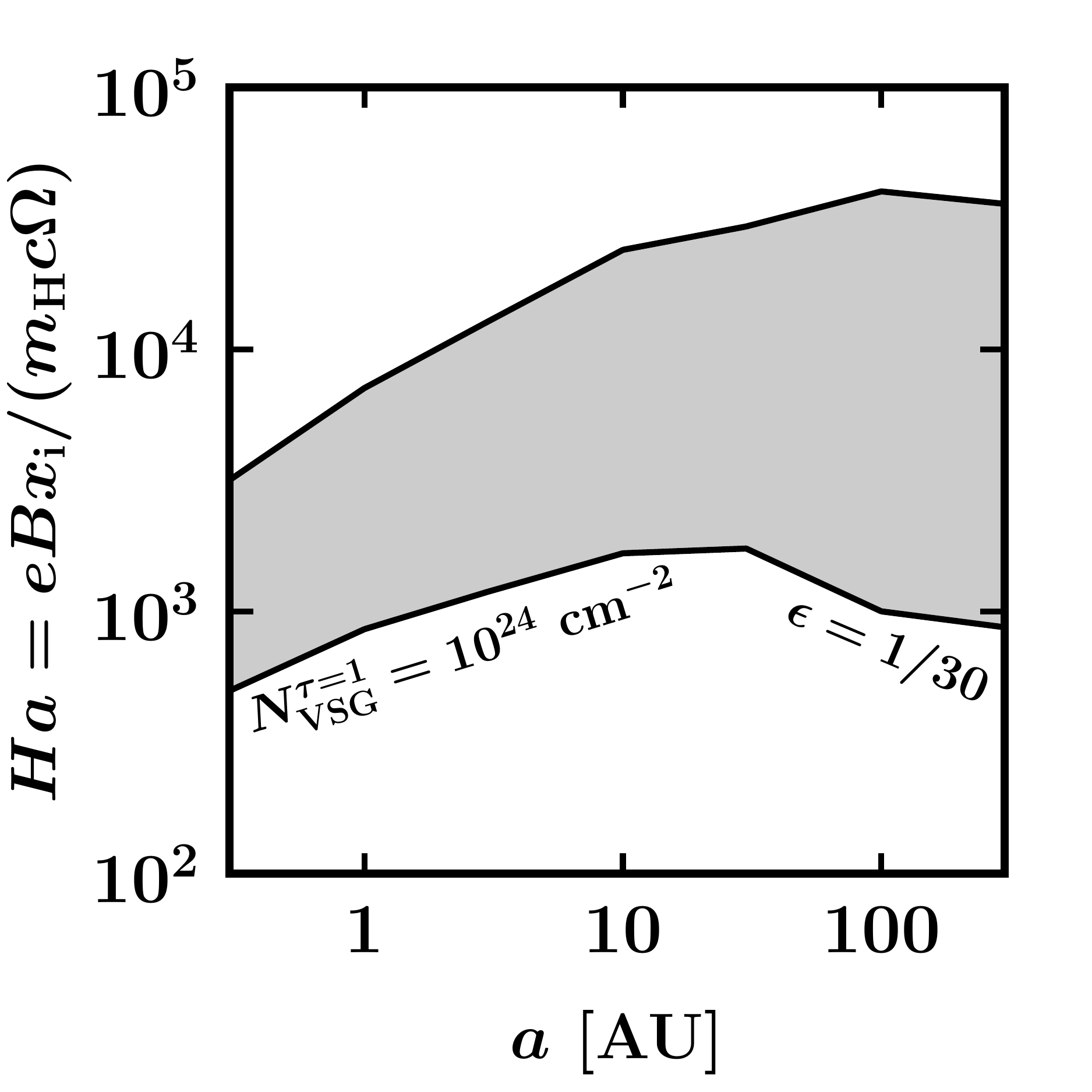}
\caption{Hall parameter $Ha$ as a function of $a$ in FUV-ionized surface layers. Values of $B$ are computing using equations (\ref{schwarz}) and (\ref{stress}).  The upper bounding
  curve is computed for our standard (fiducial) parameters, at
  $\Sigma^\ast \approx 0.01$ g/cm$^{2}$. The lower bounding curve is
  computed for a model depleted in dust, carbon, and sulfur, for
  which $\Sigma^\ast \approx 0.2$ g/cm$^{2}$ (see also Figure
  \ref{fig_sensitivity}a). Hall diffusion is not significant for FUV-irradiated
surface layers.}
\label{fig_hall}
\end{figure}

When computing $\Sigma^*$, we have ignored the effects of Hall
diffusion \citep{Wardle:1999p11536, Balbus:2001p11539, Sano:2002p7863}.
When Hall diffusion dominates, only electrons are coupled to magnetic
fields, and ions are de-coupled from magnetic fields by collisions
with neutrals.  Under these circumstances, the evolution of the MRI
depends on the direction of the magnetic field $\mathbf{B}$ with
respect to the angular frequency $\mathbf{\Omega}$. Hall diffusion can
increase/decrease $\Sigma^*$ by an order of magnitude or more when
$\mathbf{B}$ and $\mathbf{\Omega}$ are parallel/anti-parallel
\citep{Wardle:2011p11473}.

To assess the importance of Hall diffusion in FUV-ionized surface 
layers, we evaluate the Hall parameter
\begin{equation} \label{hall}
Ha \equiv \frac{v^2_A}{\eta_{Ha}\Omega} = \frac{e B x\sub{i}}{ m\sub{H} c \Omega} \,.
\end{equation}
Here $x\sub{i} \equiv n\sub{i}/n\sub{tot}$ is the fractional abundance
of ions of density $n\sub{i}$ relative to hydrogen nuclei of density
$n\sub{tot}$;
$v_A = B/\sqrt{4\pi n\sub{tot}m\sub{H}}$ is the Alfv\'en velocity; $e$ is
the electron charge; $c$ is the speed of light; and
$\eta_{Ha}= c B/(4\pi e n\sub{i})$ is the Hall diffusivity (e.g.,
\citealt{Wardle:2011p11473}, but note that their Hall parameter is
the inverse of ours, and they assume $n\sub{i} = n\sub{e}$).
The dimensionless Hall parameter
is the ratio of the inductive term to the Hall term in the magnetic
induction equation. If $Ha \gg 1$, then Hall diffusion is not important.

To evaluate $Ha$, we estimate $B$ from $\dot{M}$ by assuming that the inequality
\begin{equation} \label{schwarz}
B^2 \geq B_r^2 + B_\phi^2 \geq 2 B_r B_\phi 
\end{equation}
saturates, and that the Maxwell stress which drives accretion is given by
\begin{equation} \label{stress}
B_r B_\phi \approx \frac{\dot{M} \Omega}{2h}
\end{equation}
(see, e.g., \citealt{Bai:2009p3370}; the factor of 2 in equation \ref{stress}
arises because the disk has a top and bottom face).

In Figure \ref{fig_hall} we show that $Ha \gg 1$ at $\Sigma^*$ for all
$a$, over the entire parameter space that we have explored.
Thus we conclude that
Hall diffusion is not a concern in FUV-ionized surface layers (the
same may not be true for the more poorly ionized X-ray-irradiated layers;
\citealt{Wardle:2011p11473}).

\subsection{The Possibility of Turbulent Mixing: Chemical Equilibration Timescale vs. Dynamical Timescale} \label{sec_trec}

\begin{figure} %
\epsscale{1.0}
\plotone{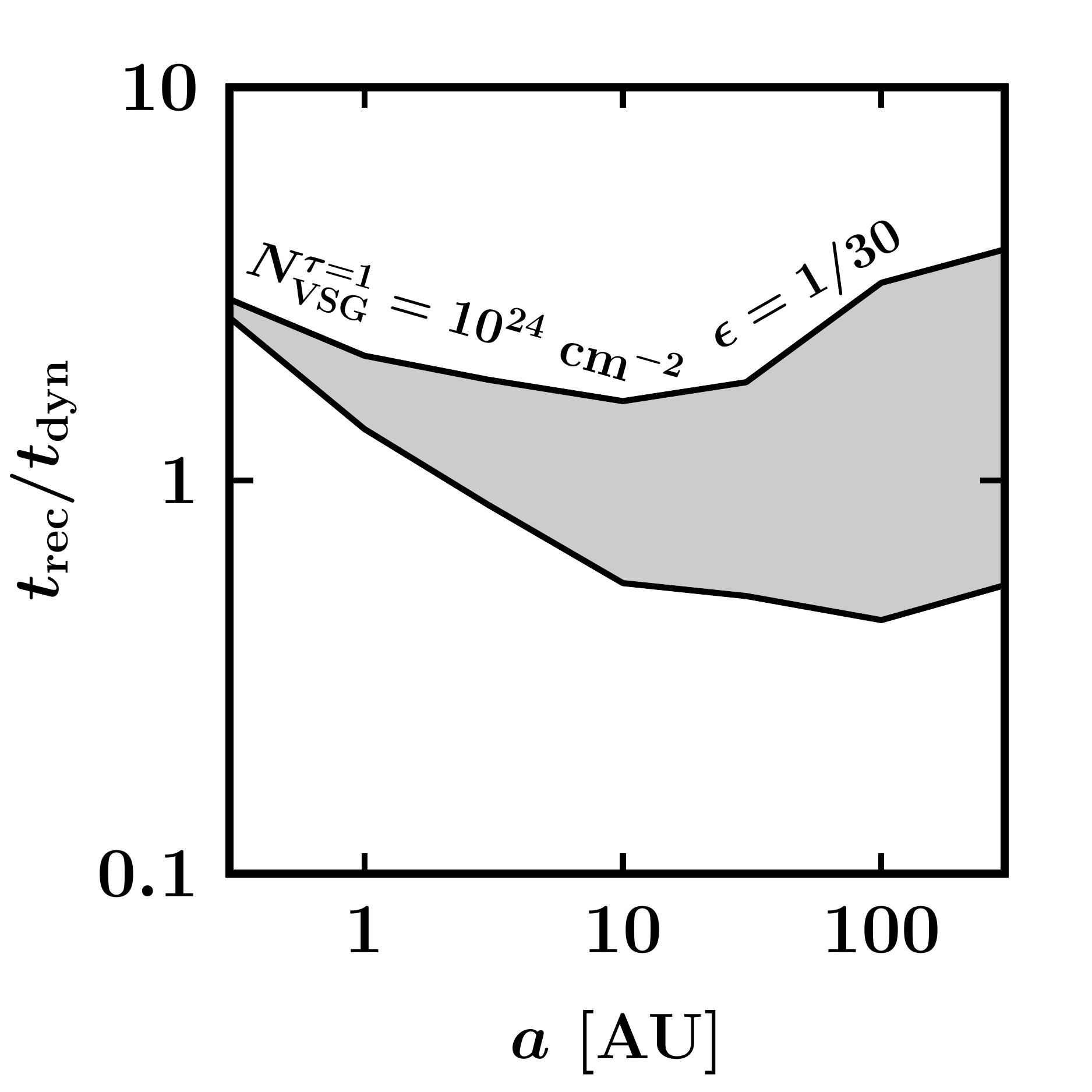}
\caption{Ratio of the ion-electron recombination timescale
  $t\sub{rec}=(n\sub{e} \alpha\sub{rec,S})^{-1}$ to the dynamical time
  $t\sub{dyn}=\Omega^{-1}$, as a function of $a$. The lower bounding
  curve is computed for our standard (fiducial) parameters, at
  $\Sigma^\ast \approx 0.01$ g/cm$^{2}$. The upper bounding curve is
  computed for a dust-depleted and carbon/sulfur-depleted model, for
  which $\Sigma^\ast \approx 0.2$ g/cm$^{2}$ (see also Figure
  \ref{fig_sensitivity}).}
\label{fig_TrecFUV}
\end{figure}

Throughout this paper, we have computed ionization fractions in a
static atmosphere. But the MRI-active layer is not static; it is
turbulent. We might have underestimated $\Sigma^\ast$ because
turbulence can mix plasma vertically toward the midplane, deeper into the disk
interior \citep{Inutsuka:2005p11464,Ilgner:2006p10013,Turner:2007p3503}.
Mixing would be effective
if the vertical mixing time is short compared to the timescale over
which ionized layers achieve chemical equilibrium. Without a
full-out simulation of MRI turbulence, we approximate the vertical
mixing time as the dynamical time $t_{\rm dyn} = \Omega^{-1}$.  For
the chemical equilibration time, we substitute the radiative
recombination time $t_{\rm rec} = 1/(n\sub{e} \alpha\sub{rec,S})$. The
ratio $t_{\rm rec}/t_{\rm dyn}$ is plotted in Figure \ref{fig_TrecFUV}
(cf. Figure 11 of PBC11). Because $t_{\rm rec}/t_{\rm dyn} > 1$ over
some portion of parameter space, mixing might well be significant in
FUV-ionized layers, and might increase $\Sigma^\ast$ above the values we
have computed in this paper.  Whether the increase would be large or small is hard
to say. On the one hand, $t_{\rm rec}/t_{\rm dyn}$ is never far above
unity, suggesting that turbulent mixing will merely introduce
order-unity corrections to our estimates. On the other hand, as turbulent mixing
dilutes the electron density $n\sub{e}$, the timescale ratio $t_{\rm
  rec}/t_{\rm dyn}$ might increase with increasing depth, and the
process might run away.

\section{SUMMARY AND IMPLICATIONS FOR DISK
  ACCRETION} \label{discussion}

Circumstellar disk material must be sufficiently ionized if it is to
accrete by the magnetorotational instability (MRI).  We have
considered in this paper ionization by stellar far-ultraviolet (FUV)
radiation. Although FUV radiation cannot penetrate the disk as deeply
as can X-rays, it generates ionization fractions orders of magnitude
larger.  Carbon and sulfur may be the principal sources of free
electrons and ions, as these elements are cosmically abundant and
least likely to be depleted onto grains.  Far-infrared searches for
carbon and sulfur emission from disks are so far consistent with gas
phase abundances for both elements within a factor of $\sim$2 of
solar. Because the electron and ion abundances generated from ionized
carbon and sulfur are so large, ionization fractions in FUV-irradiated
layers are little impacted by PAHs. In FUV-ionized layers, fractional
ion abundances $x_{\rm i} \approx 10^{-5}$--$10^{-4}$, dwarfing PAH
abundances of $x_{\rm PAH} \sim 10^{-11}$--$10^{-8}$. More to the point,
ion recombination on PAHs in FUV-ionized layers
is at most competitive with ion recombination with free electrons.
This immunity to
PAHs does not apply to X-ray ionized layers where $x_{\rm i} \lesssim
10^{-9}$ and where PAHs or very small (0.01 $\mu$m sized) grains can suppress the MRI
(\citealt{Bai:2009p3370}; \citealt{PerezBecker:2011p9801}, PBC11).
It is refreshing that FUV ionization is robust against the usual
difficulties plaguing other, weaker sources of ionization.

\begin{figure} %
\epsscale{1.0}
\plotone{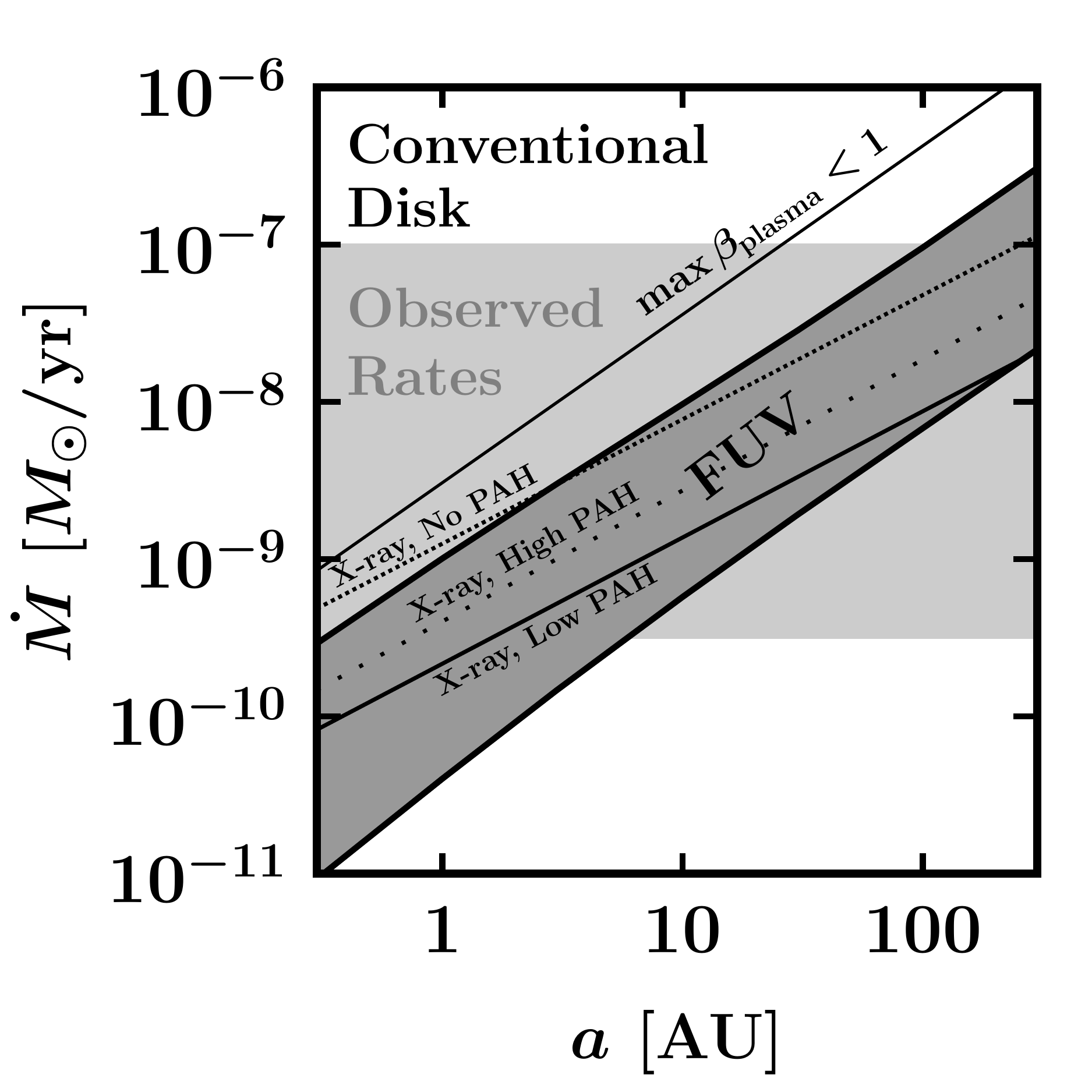}
\caption{
  Accretion rates for conventional (hole-less) disks. The
  shaded region labeled ``FUV'' (bounded by heavy solid lines)
  corresponds to the range of accretion
  rates possibly driven by the MRI in FUV-ionized surface layers. The
  range of FUV-driven accretion rates corresponds to the range of
  MRI-active surface densities $\Sigma^\ast$ shown in Figure
  \ref{fig_sensitivity}a, which in turn corresponds to a range of
  possible abundances for grains, carbon, and sulfur (PAHs, whose abundances
  for all our FUV models are set to their maximum value, are not significant
  for FUV-ionized layers). 
  We compute X-ray-driven accretion rates using our corrected
definition for $Am$ (Equations \ref{Am}--\ref{ncharge}), for the
low-PAH and high-PAH cases considered by PBC11. The case with no PAHs
is shown for comparison only.
 Note how little the X-ray-driven $\dot{M}$ varies, despite the
PAH abundance having changed by three orders of magnitude; the small
variation in $\dot{M}$ follows from Equation (\ref{ncharge}), in which
reductions in $n\sub{i}$ from charge-neutralizing PAHs are offset by
the increased abundances of charged PAHs themselves. In fact, the
high-PAH $\dot{M}$ even exceeds slightly the low-PAH $\dot{M}$ (see
also \citealt{Bai:2011p0001}).
At $a \lesssim 10$ AU, accretion rates
  in FUV-ionized layers may be comparable to those in X-ray ionized
  layers.  At $a \gtrsim 10$ AU, FUV-driven accretion rates tend to be
  larger than X-ray-driven rates.  The light shaded region labeled
  ``Observed Rates'' brackets the range of stellar accretion rates for
  stars of mass $M \approx 0.3$--$1 M_{\odot}$ as shown in Figure 5 of
  \citet{Muzerolle:2005p9975}.}
\label{fig_mdotxfuv}
\end{figure}

In Figure \ref{fig_mdotxfuv} we compute the possible ranges of disk
accretion rate $\dot{M}$ due to either FUV or X-ray ionization, and
compare them against the range of observed stellar accretion rates.
The accretion rate $\dot{M}$ driven by FUV ionization is derived using
equations (\ref{maxalpha}) and (\ref{mdot_define}) for $\Sigma =
\Sigma^\ast$ (taken from Figure \ref{fig_sensitivity}a) and the
corresponding value of $\max \alpha$.  The accretion rate $\dot{M}$
driven by X-ray ionization is computed similarly, using the abundances
of charged species of the standard model of PBC11, their
temperature of $T = 80 (a/{\rm 3 AU})^{-3/7}$ K, and Equations
(\ref{Am})--(\ref{ncharge}) of this work to compute $Am(\Sigma)$.
Although our estimates of $\dot{M}$ driven by X-ray ionization
may have to be revised because of the Hall effect (\citealt{Wardle:2011p11473}),
our estimates of $\dot{M}$ driven by FUV ionization should not be: FUV-ionized,
MRI-active layers behave in the ideal magnetohydrodynamic limit.

\begin{figure} %
\epsscale{1.0}
\plotone{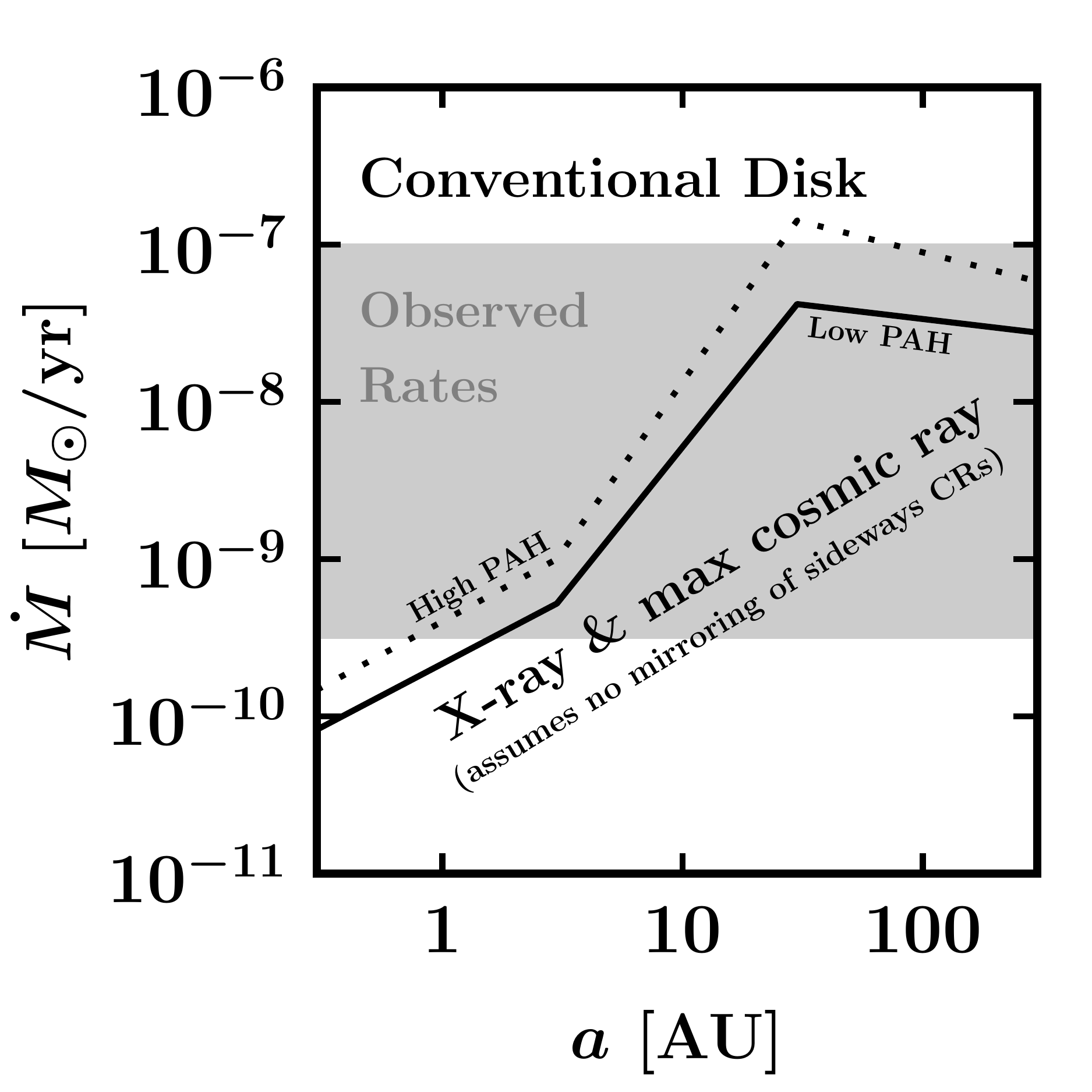}
\caption{Same as Figure \ref{fig_mdotxfuv}, but showing disk accretion
  rates driven by the combined effects of X-ray and cosmic ray
  ionization. Galactic cosmic rays cannot reach the disk through its
  top and bottom faces because of shielding by the magnetized stellar
  wind. They can, however, penetrate the disk ``sideways'' through the
  outermost portions of the disk, parallel to the midplane---assuming they
  are not magnetically mirrored away. 
  To generate this plot we assume a total (vertical) surface density
  of $2200 (a/1 \, {\rm AU})^{-3/2}$ g/cm$^2$, and a cosmic ray
  ionization rate $\zeta_{0} \exp(-\Sigma_{a}/\Sigma_0$), where
  $\zeta_{0}= 1/4 \times 10^{-17}$ s$^{-1}$ \citep{Caselli:1998p3151}
  and $\Sigma_0 = 96$ g/cm$^{2}$ is the stopping column for GeV-energy
  cosmic rays \citep{Umebayashi:1981p7866}. The factor of 1/4 in
  $\zeta_0$ is our estimate for the fraction of the celestial sphere
  (centered at the midplane) not shielded by stellar winds. The
  attenuating column $\Sigma_{a}$ is obtained by radially integrating
  the volume mass density at the midplane from $a$ to infinity. For $a
  \gtrsim 30$ AU, sideways cosmic rays may provide enough ionization
  to render the entire disk MRI active (PBC11).  Sideways cosmic rays
  cannot penetrate the disk at $a \lesssim 10$ AU because of
  intervening disk material, and so the accretion rates there are
  practically identical to those for the X-ray-only case (plotted in Figure
  \ref{fig_mdotxfuv}).  The solid and dotted curves correspond to the cases of low and high PAH abundances, respectively. The main assumption underlying this plot is that sideways cosmic rays are not magnetically mirrored away; this is an uncertain prospect.}
\label{fig_mdotcr}
\end{figure}

Modulo the impact of the Hall effect on X-ray-driven MRI, from
Figure \ref{fig_mdotxfuv} we conclude that accretion rates from
FUV-ionized surface layers are of the same order of magnitude as
accretion rates from X-ray ionized layers. At large radii $a \gtrsim
10$ AU, FUV-driven accretion rates may exceed X-ray-driven rates. At
$a \lesssim 10$ AU, their contributions may be more nearly equal.
At $a \gtrsim 10$ AU, the FUV-irradiated surface layer can
sustain accretion rates similar to those observed.  At $a \lesssim 10$
AU, surface layer accretion rates, driven either by FUV or X-ray ionization, can still be observationally
significant, but they diminish with decreasing radius.  At $a \lesssim
1$ AU, surface layer accretion rates fall below the range typically
observed for young solar-type stars.

The problem of too low an accretion rate at small radius might be
alleviated by turbulent mixing of plasma from FUV-irradiated
disk surface layers into
the disk interior, as such mixing would enhance the thickness of the MRI-active
layer.  Our crude estimate of the timescales involved (Section
\ref{sec_trec}) suggests that this possibility is worth further consideration.

Note in Figure 8 how the X-ray-driven $\dot{M}$ actually
increases from the low-PAH case to the high-PAH case. This surprising effect
arises because increasing the PAH abundance (over this particular range from
$x_{\rm PAH} = 10^{-11}$ to $10^{-8}$) increases the abundances of charged
PAHs themselves, which more than offsets the decreased abundances of free
electrons and ions. In fact, in the high-PAH case, positively and negatively
charged PAHs are approximately equal in number and represent by far the most
abundant charged species. The primary recombination pathway is positively
charged PAHs colliding with negatively charged PAHs (reaction 14 in PBC11).
Thus as the overall PAH abundance increases, charged PAHs can increase $Am$ at
a given column $\Sigma$. This effect was missed by PBC11, and is highlighted
by \citet{Bai:2011p0001}. The consequent enhancement in $\dot{M}$ is mitigated
by the decrease in the free electron fraction and thus in $Re$. In computing
the X-ray-driven $\dot{M}$ for the high-PAH case in Figure 8, we accounted for
the mitigating effects of a lower $Re$ by evaluating $Am$ at the $\Sigma$ for
which $Re \approx 100$, below which Ohmic dissipation would weaken the MRI.

One concern is whether gas pressures in FUV-ionized layers
are lower than the magnetic field pressures required to drive
our computed accretion rates. If the pressure ratio
\begin{equation} \label{plasmabeta}
\beta_{\rm plasma} \equiv \frac{\Sigma^\ast k T / (\mu h)}{B^2/8\pi}
\end{equation}
is $< 1$, magnetic tension defeats the MRI. An upper bound on $\beta_{\rm plasma}$ is obtained by substituting the lower bound on $B^2$ inferred from
equations (\ref{schwarz}) and (\ref{stress}).
For a representative
case $\Sigma^\ast = 0.1$
g/cm$^2$, equations (\ref{schwarz})--(\ref{plasmabeta}) combine to
yield the topmost slanted line in Figure \ref{fig_mdotxfuv}, above
which $\max \beta_{\rm plasma} < 1$ and the MRI cannot operate. The
range of FUV-driven accretion rates we have computed sits safely below
this line.

In sum, surface layer accretion by FUV ionization can, by itself,
solve the problem of protoplanetary disk accretion at large radius,
but not at small radius (unless turbulent mixing of plasma can
substantially thicken the MRI-active layer). This statement remains
one of principle and not of fact, because MRI accretion rates depend
on the strength and geometry of the background magnetic field
threading the disk (e.g., \citealt{Fleming:2000p6431};
\citealt{Pessah:2007p10630}; Bai \& Stone 2011), and unfortunately the
field parameters are not known for actual disks.\footnote{The rotation
  measures of magnetized FUV-ionized layers could be large, on the
  order of $10^4$ rad/m$^2$ at $a \sim 30$ AU for a face-on disk
  (assuming an untangled $B \sim 10 $mG and an electron column density
  of $N\sub{e} \sim 3\times10^{18}$ cm$^{-2}$).  Perhaps measurements
  of Faraday rotation at radio wavelengths using polarized background
  active galaxies, or polarized emission from the central star itself,
  could be used to constrain disk magnetic fields.}

For completeness, we show in Figure \ref{fig_mdotcr} how sideways
cosmic rays can, in principle, enhance accretion rates in the
outermost portions of disks (see the Introduction). For an assumed
radial surface density profile resembling that of the minimum-mass
solar nebula, accretion rates at $a \gtrsim 30$ AU can be strongly
increased by sideways cosmic rays.  Although cosmic-ray induced
accretion rates at large radius can be competitive with FUV-induced
rates, it is not clear that cosmic rays are not mirrored away from
disks by ambient magnetic fields.  Far-UV ionization does not suffer
from this uncertainty.

\subsection{Transitional Disks}

\begin{figure} %
\epsscale{1.0}
\plotone{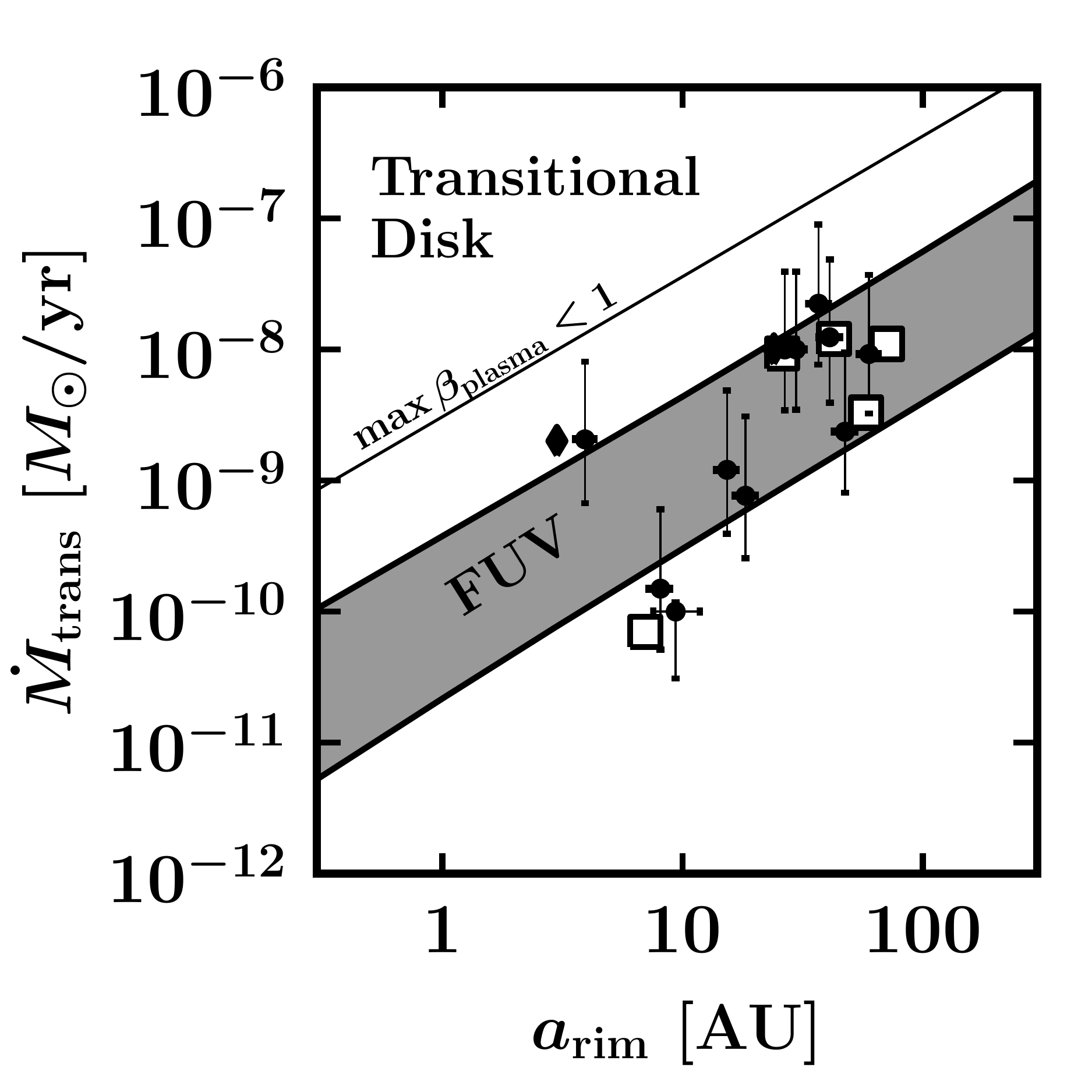}
\caption{Same as Figure \ref{fig_mdotxfuv} but for transitional
  disks and showing only our FUV model. Observed rates are taken from
  \citet[][diamonds]{Calvet:2005p2200},
  \citet[][squares]{Espaillat:2007p5236,Espaillat:2008p11248,Espaillat:2010p7337},
and
  \citet[][circles with error bars]{Kim:2009p5659}. Theoretical rates
  for FUV-ionized layers are computed with equation
  (\ref{mdot_trans}), with $\Sigma^\ast$ recomputed using $\theta \sim
  1$.  The observed trend of increasing $\dot{M}_{\rm trans}$ with
  increasing $a_{\rm rim}$ is reproduced 
(see, however, Section 4.2 of \citealt{Kim:2009p5659} for how the
  observed trend could reflect other correlations between stellar mass $M$ and $a_{\rm rim}$, and $M$ and $\dot{M}$).
}
\label{fig_mdotxfuv-trans}
\end{figure}

What is the relevance of MRI accretion to transitional disks, i.e.,
disks with inner holes (e.g., \citealt{Espaillat:2010p7337};
\citealt{Hughes:2007p2254}; \citealt{Calvet:2005p2200})?
\citeauthor{Chiang:2007p3804} (2007, hereafter CMC) proposed that
MRI-active surface layers at the rim of the hole---i.e., in the rim
``wall'' oriented perpendicular to the disk midplane---could supply
the observed accretion rates of transitional disks. Following CMC, we
compute the accretion rate at the rim according to the formula
\begin{equation} \label{mdot_trans}
\dot{M}_{\rm trans} \sim \frac{3 M_{\rm rim}}{t_{\rm diff}} \sim 
12\pi \Sigma^\ast \times \max \alpha \times \left( \frac{kT}{\mu} \right)^{3/2} \frac{a_{\rm rim}^2}{GM}
\end{equation}
where $M_{\rm rim} \approx 4 \pi a h \Sigma^\ast$ ($h$ is the disk
half-thickness), $t_{\rm diff} \sim a_{\rm rim}^2/\nu$ is the time for
material to diffuse from $a_{\rm rim}$ to $a_{\rm rim}/2$, and
$\Sigma^\ast$ is reinterpreted for transitional disk rims as the
radial, not vertical, column of MRI-active material.
We re-compute $\Sigma^\ast$ for transitional
disk rims by taking the grazing angle
$\theta \sim 1$, as is appropriate for radiation
which penetrates the rim wall at normal incidence.

Figure \ref{fig_mdotxfuv-trans} shows accretion rates for transitional
disk rims computed according to (\ref{mdot_trans}), overlaid with
observations taken from \citet{Calvet:2005p2200}, \citet{Espaillat:2007p5236,Espaillat:2008p11248,Espaillat:2010p7337}, and \citet{Kim:2009p5659}. We
have chosen these references and not others because
they utilize disk models similar enough to each other to yield
reasonably consistent hole radii (cf. \citealt{Merin:2010p11143}).
Rim accretion by FUV
ionization seems capable of reproducing the trend of increasing
$\dot{M}_{\rm trans}$ with increasing $a_{\rm rim}$. However, there
remains the problem of transporting the material that is dislodged
from the rim over the decades in disk radius between the rim and the
host star.  Inside the rim, a conventional disk geometry applies,
and we have already noted in that context that surface layer accretion
rates at small radii are too small compared to those observed.  For
example, according to Figure \ref{fig_mdotxfuv-trans}, an accretion
rate of $\dot{M}_{\rm trans} \sim 10^{-9} M_\odot/$yr may be initiated
at $a_{\rm rim} =$ 30 AU, but according to Figure \ref{fig_mdotxfuv},
this same accretion rate cannot be sustained by the MRI inside a
radius of $\sim$3 AU. Multiple planets could solve this problem by
shuttling gas inward (PBC11; \citealt{Zhu:2011p11465}).

\acknowledgements We thank M\'at\'e \'Ad\'amkovics, Xue-Ning Bai,
Barbara Ercolano, Al Glassgold, Carl Heiles, Meredith Hughes, Subu
Mohanty, Ruth Murray-Clay, James Owen, Dima Semenov, and Neal Turner
for discussions. Xue-Ning Bai has generously shared advance copies of his
papers on X-ray-driven MRI that have significantly
informed our work. 
We are grateful to Greg Herczeg and Roy van Boekel
for organizing a Ringberg conference that led to many refinements in
our analysis.  We especially would like to thank the anonymous referee
whose careful reading of this paper and PBC11 motivated several
improvements in the presentation of our results.  This work was
supported by a National Science Foundation Graduate Research
Fellowship awarded to D.P.-B.

\end{document}